\documentclass[aps,
               pra,
               twocolumn,
               showpacs,
               amsmath,
               amssymb,
               reprint,
               10pt
              ]{revtex4-1}

\usepackage[hyperindex,colorlinks,breaklinks,allcolors=blue]{hyperref}
\usepackage{txfonts}

\usepackage{amsmath, amssymb}

\usepackage{pstricks}
\usepackage{graphicx}
\usepackage{tikz}
\usepackage[utf8x]{inputenc}
\usepackage{color}

\usepackage{pstool}
\usepackage{pgf}
\usepackage{import}

\usetikzlibrary{snakes}
\usetikzlibrary{arrows}

\usepackage{tabularx}
\usepackage{multirow}
\usepackage{array}
\newcolumntype{L}[1]{>{\raggedright\let\newline\\\arraybackslash\hspace{0pt}}m{#1}}
\newcolumntype{C}[1]{>{\centering\let\newline\\\arraybackslash\hspace{0pt}}m{#1}}
\newcolumntype{R}[1]{>{\raggedleft\let\newline\\\arraybackslash\hspace{0pt}}m{#1}}

\makeatletter
\let\cat@comma@active\@empty
\makeatother

\begin{document}
\title{Non-thermal fixed points: Universal dynamics far from equilibrium}
\thanks{This overview article has been written for the proceedings of the \href{http://www.schwinger100.org}{\emph{Julian Schwinger Centennial Conference and Workshop}} held in Singapore, 7 to 12 February 2018.}

\author{Christian-Marcel Schmied}
\email{christian-marcel.schmied@kip.uni-heidelberg.de}

\author{Aleksandr N. Mikheev}
\email{aleksandr.mikheev@kip.uni-heidelberg.de}

\author{Thomas~Gasenzer}
\email{t.gasenzer@uni-heidelberg.de}
\affiliation{Kirchhoff-Institut f\"ur Physik, 
             Ruprecht-Karls-Universit\"at Heidelberg,
             Im Neuenheimer Feld 227, 
             69120 Heidelberg, Germany}

\date{\today}

\begin{abstract}
In this article we give an overview of the concept of universal dynamics near non-thermal fixed points in isolated quantum many-body systems.
We outline a non-perturbative kinetic theory derived within a Schwinger-Keldysh closed-time path-integral approach, as well as a low-energy effective field theory which enable us to predict the universal scaling
exponents characterizing the time evolution at the fixed point. 
We discuss the role of wave-turbulent transport in the context of such fixed points and discuss universal scaling evolution of systems bearing ensembles of (quasi) topological defects. 
This is rounded off by the recently introduced concept of prescaling as a generic feature of the evolution towards a non-thermal fixed point. 
\end{abstract}


\maketitle

\section{Introduction}
\label{sec:Intro}
Dynamics of isolated quantum many-body systems quenched far from equilibrium has been an object of intensive study during recent years. 
Examples range from the post-inflationary early universe \cite{Kofman:1994rk,Micha:2002ey}, via the dynamics of quark-gluon matter created in heavy-ion collisions \cite{Baier:2000sb,Berges:2013eia} to the evolution of ultracold atomic systems following a sudden quench of, e.g., an interaction parameter \cite{Bloch2008a.RevModPhys.80.885,Polkovnikov2011a.RevModPhys.83.863}. 
Yet, despite great efforts, there are many open questions remaining and little is known about the structure of possible pathways for the evolution of such systems.
Various scenarios have been discussed for and observed in ultracold atomic gases, including 
integrable dynamics~\cite{Kinoshita2006a,Hofferberth2007a.2007Natur.449..324H,Trotzky2012a.NatPhys8.325,Calabrese2016b}, 
prethermalization~\cite{Bettencourt:1997nf,Aarts2000a.PhysRevD.63.025012,Berges:2004ce,Gring2011a,Langen:2016vdb}, 
generalized Gibbs ensembles (GGE)~\cite{Jaynes1957a, Jaynes1957b,Rigol2007a.PhysRevLett.98.050405,Langen2015b.Science348.207,Vidmar2016JSMTE..06.4007V}, 
critical and prethermal dynamics~\cite{Braun2014a.arXiv1403.7199B,Nicklas:2015gwa,Navon2015a.Science.347.167N,Eigen2018a.arXiv180509802E}, 
many-body localization~\cite{Schreiber2015a.Science349.842,Vasseur2016a.160306618V}, 
relaxation after quantum quenches \cite{Essler2016a.JSMTE..06.4002E,Cazalilla2016a.160304252C}, 
wave turbulence \cite{Zakharov1992a, Nazarenko2011a,Navon2016a.Nature.539.72}, 
decoherence and revivals~\cite{Rauer2017a.arXiv170508231R.Science360.307}, 
universal scaling dynamics and the approach of a non-thermal fixed point~\cite{Berges:2008wm,Orioli:2015dxa,Prufer:2018hto,Erne:2018gmz}, 
as well as prescaling~\cite{Schmied:2018upn.PhysRevLett.122.170404}.
The rich spectrum of different possible phenomena highlights the capabilities of experiments with ultracold gases, 
as well as the gain obtained with quantum systems as compared to classical statistical ensembles.

To theoretically study such out-of-equilibrium phenomena in quantum many-body systems, a wide range of tools and techniques provided by nonequilibrium quantum field theory is used.
A central object in calculating the non-equilibrium time evolution of a many-body system is the so-called Schwinger-Keldysh closed time contour. 
Evolving quantities such as correlation functions of physical observables in time corresponds to evaluating, e.g., path integrals along such a closed contour.
The technique was first introduced by Julian Schwinger in 1961~\cite{Schwinger1961a} and further developed by Mahanthappa and Bakshi
\cite{Mahanthappa1961a.PhysRev.126.329,Bakshi1963a,Bakshi1963b}, who were focussing on bosonic systems.
Around the same time, Konstantinov and Perel developed a diagrammatic scheme for evaluating transport quantities in nonequilibrium systems.
They used a time contour with forward and backward branches in time together with an imaginary-time path whose length is given by the inverse temperature \cite{Konstantinov1960}. 
The framework of nonequilibrium quantum field theory was then advanced by Kadanoff and Baym in 1962 \cite{KadanoffBaym1962a}. 
In their work, they also showed a pathway to kinetic equations. 
Keldysh proposed the closed-time path technique in 1964 and introduced a convenient choice of variables via the so-called Keldysh rotation \cite{Keldysh1964a}. 
To acknowledge all this work, the closed time contour is referred to as the Schwinger-Bakshi-Mahanthappa-Keldysh, in short Schwinger-Keldysh formulation \footnote{Cf.~Ref.~\cite{Mihaila:2000sr}, using a different order of the names. See Refs.~\cite{Keldysh2002a,Kamenev2011} for more detailed historical notes. Taking into account the close connection to the Kadanoff-Baym kinetic equations near equilibrium, including an imaginary-time branch, the approach has also been referred to as the Baym-Kadanoff-Schwinger-Mahanthappa-Bakshi-Keldysh formalism \cite{Martin1999a}.}.        
The fermionic case was later considered by Larkin and Ovchinnikov in the context of superconductivity \cite{Larkin1975a}.
In summary, based on the work of Schwinger, it became possible to describe a wealth of non-equilibrium phenomena in quantum many-body systems, including non-thermal fixed points which are the subject of the present contribution.

Our article is organized as follows. 
In Sec.~\ref{sec:NTFP} we introduce the basic concepts of non-thermal fixed points for which we outline, in Sec.~\ref{sec:KineticTheory}, a non-perturbative kinetic-theory description based on the Schwinger-Keldysh contour. 
A complementary approach in the form of a low-energy effective field theory description is presented in Sec.~\ref{sec:LEEFT}.
We address the relation between (wave) turbulence and non-thermal fixed points  in Sec.~\ref{sec:Turbulence} and
compare the analytical predictions with numerical simulations in Sec.~\ref{sec:Numerics}.
We finally discuss,  in Sec.~\ref{sec:Prescaling}, the recently proposed concept of prescaling as a generic feature of the evolution towards a non-thermal fixed point.
Our brief overview, which tries to give a short introduction without aiming at a full review, closes with an outlook to future research in the field, see Sec.~\ref{sec:Outlook}.

\section{Non-thermal fixed points}
\label{sec:NTFP}
\begin{figure}[t]%
\begin{center}
  \includegraphics[width=0.97\columnwidth]{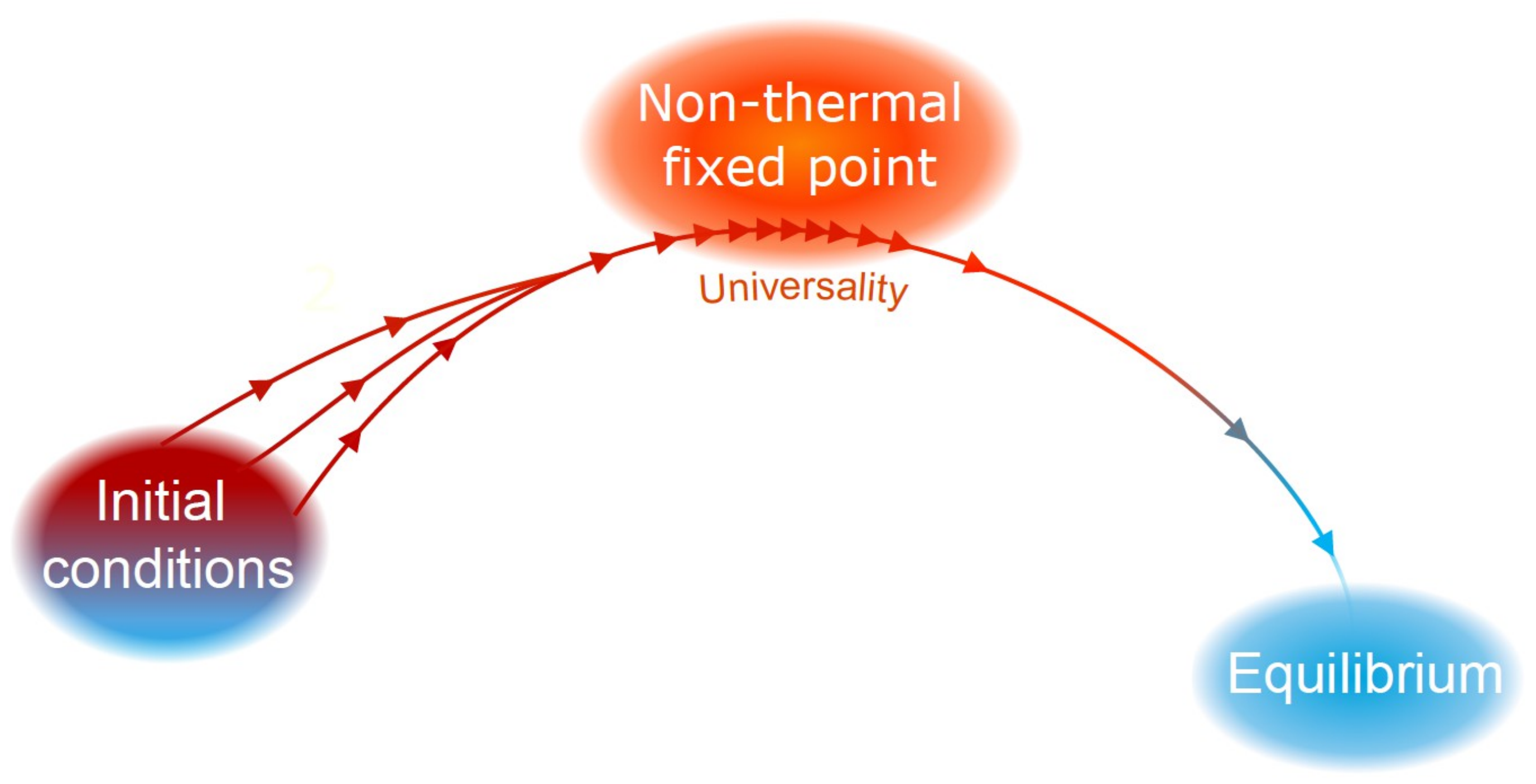}
  \caption{Schematics of a non-thermal fixed point \cite{Berges:2008wm} based on the ideas of a renormalization group flow.
  Depending on the initial condition, an out-of-equilibrium system can approach a non-thermal fixed point during the time evolution.
  In the vicinity of such a fixed point, the system experiences critical slowing down (indicated by the tightly packed red arrows). As a consequence, correlation functions $C(k,t)$ show scaling behavior in space and time according to $C(k,t)=t^{\alpha}f(t^{\beta}k)$, with a universal scaling function $f$.
The associated self-similar evolution is, in general, characterized by non-zero universal scaling exponents $\alpha$ and $\beta$.
Universal scaling close to a non-thermal fixed point is understood to occur as a transient phenomenon on the way to equilibrium (indicated by the trajectory leading away from the fixed point). Figure adapted from Ref.~\cite{Prufer:2018hto}.
}%
  \label{fig:ntfp-flow}
\end{center}
\end{figure}

The theory of non-thermal fixed points in the real-time evolution of, foremost closed, nonequilibrium systems, is inspired by the concepts of equilibrium and near-equilibrium renormalization-group theory \cite{Hohenberg1977a,Goldenfeld1992a,ZinnJustin2004a}, see Fig.~\ref{fig:ntfp-flow} for an illustration. 

The basic concept is motivated by universal critical scaling of correlation functions in equilibrium.
When using a renormalization-group approach, a physical system is basically studied through a microscope at different resolutions.
Close to a phase transition one observes that the correlations look self-similar, i.e., the same no matter which resolution is used.
In this case, shifting the spatial resolution by a multiplicative scale parameter $s$ causes correlations between points with distance $x$, denoted by $C(x;s)$, to be rescaled according to $C(x;s)=s^{\zeta}f(x/s)$. 
Hence, the correlations are solely characterized by a universal exponent $\zeta$ and the universal scaling function $f$.
A fixed point of the renormalization-group flow is reached when a change of the scale $s$ does not change $C$ by any means.
In that case the scaling function takes the form of a pure power law $f(x) \sim x^{-\zeta}$.
In a realistic physical system, the scaling function $f$ is, in general, not a pure power law but retains information of characteristic scales such as a correlation length $\xi$.
Thus, the system may only approximately reach the fixed point.

Taking the time $t$ as the scale parameter, the renormalization-group idea can be extended to time evolution of systems (far) away from equilibrium. 
The corresponding fixed point of the renormalization-group flow is called non-thermal fixed point. 
In the scaling regime near a non-thermal fixed point, the evolution of, e.g., the time-dependent version of the correlations discussed above is determined by $C(x,t)=t^{\alpha}f(t^{-\beta}x)$, with two universal exponents $\alpha$ and $\beta$ which assume, in general, non-zero values. 
The associated correlation length of the system changes as a power of time, $\xi(t)\sim t^{\beta}$.
Note that the time evolution taking power-law characteristics is equivalent to critical slowing down, here in real time.
We remark that, depending on the sign of $\beta$, increasing the time $t$ can  correspond to either a reduction or an increase of the microscope resolution.

The scaling exponents $\alpha$ and $\beta$ together with the scaling function $f$ allow to determine the universality class associated with the fixed point \cite{Berges:2014bba,Orioli:2015dxa}.
Hence, the evolution of very different physical systems far from equilibrium can be categorized by means of their possible kinds of scaling behavior. 
While a full such classification is still lacking, 
underlying symmetries of the system are expected to be relevant for the observable universal dynamics and thus for the associated universality class.

\begin{figure}[t]
\centering
\includegraphics[width=0.95\columnwidth]{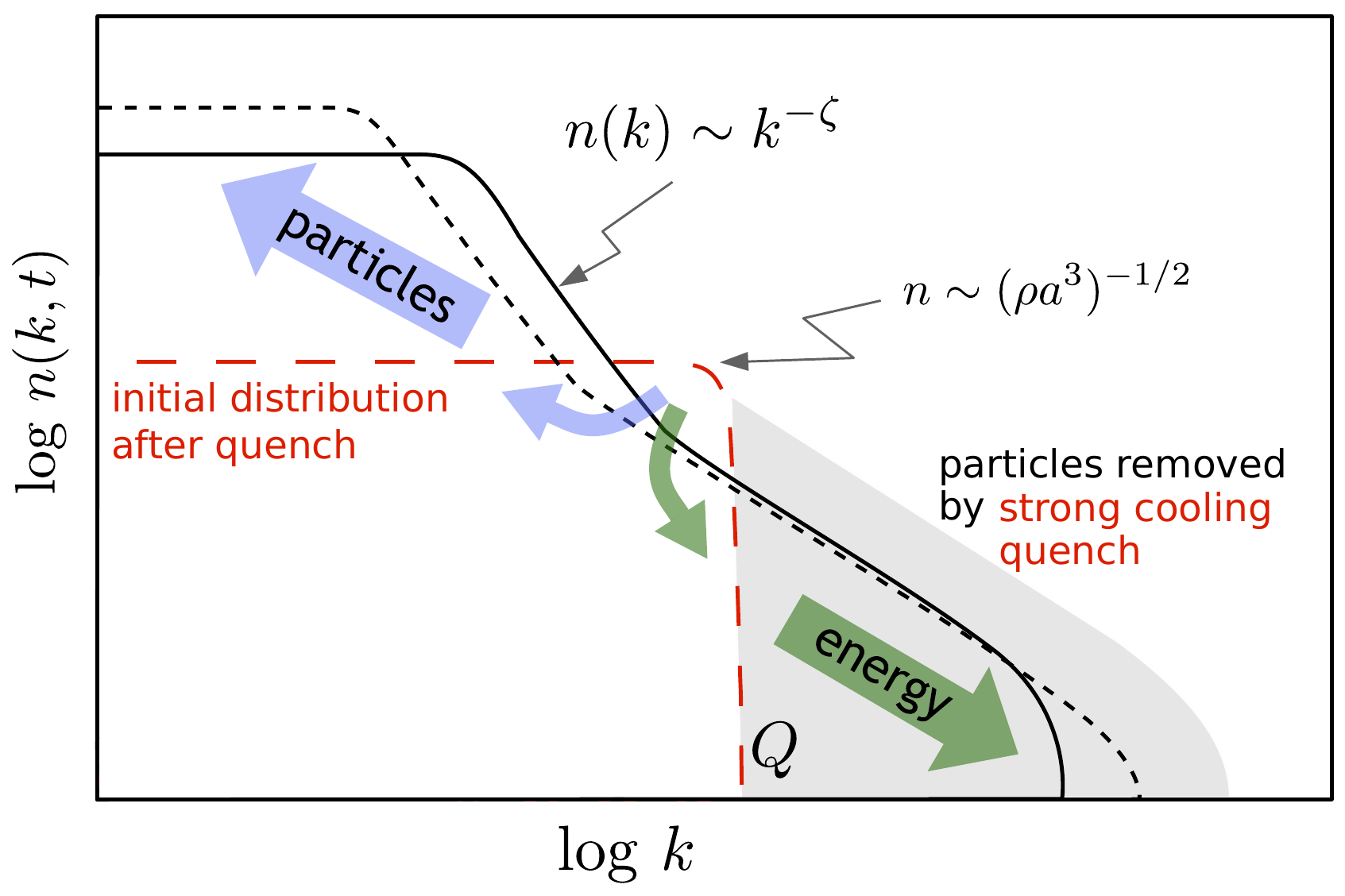}
\caption{Self-similar scaling in time and space close to a non-thermal fixed point. 
The sketch shows, on a double-logarithmic scale, the time evolution of the single-particle momentum distribution $n(k,t)$ of a Bose gas for two different times $t$ (solid and short-dashed line).  
Starting from an extreme initial distribution marked by the red long-dashed line, being the result of a strong cooling quench, a bi-directional redistribution of particles in momentum space occurs as indicated by the arrows. 
Particle transport towards zero momentum as well as energy transport to large momenta are characterized by self-similar scaling evolutions in space and time according to $n(k,t)=(t/t_{\mathrm{ref}})^{\alpha}n([t/t_{\mathrm{ref}}]^{\beta}k,t_{\mathrm{ref}})$, with universal scaling exponents $\alpha$ and $\beta$, in general, different for both directions. Here, $t_{\mathrm{ref}}$ is an arbitrary reference time within the temporal scaling regime.
The infrared transport (blue arrow) conserves the particle number which is concentrated at small momenta.
In contrast, the energy, being concentrated at high momenta, is conserved in the redistribution of short-wavelength fluctuations (green arrow).  
See main text for details.
Figure adapted from Ref.~\cite{Davis:2016hwt}.
}
\label{fig:NTFP}
\end{figure}

Whether a physical system can approach a non-thermal fixed point and show universal scaling dynamics, and, if so, which particular fixed point is reached, in general depends on the chosen initial condition. 
A key ingredient for the occurrence of self-similar dynamics is an extreme out-of-equilibrium initial configuration. 

As an illustration we consider the time evolution of a dilute Bose gas in three spatial dimensions after a strong cooling quench
\cite{Chantesana:2018qsb.PhysRevA.99.043620}, see Fig.~\ref{fig:NTFP} as well as Refs.~\cite{Nowak:2012gd,Berges:2012us,Orioli:2015dxa,Davis:2016hwt}.
An extreme version of such a quench can be achieved, e.g., by first cooling the system adiabatically such that its chemical potential is $0 < - \mu \ll k_B T$, where the temperature $T\gtrsim T_c$ is just above the critical temperature $T_c$ separating the normal and the Bose condensed phase of the gas, and then removing all particles with energy higher than $\sim \lvert \mu \rvert.$
This leads to a distribution that drops abruptly above a momentum scale $Q$ (see Fig.~\ref{fig:NTFP}).
If the corresponding energy is on the order of the ground-state energy of the post-quench fully condensed gas, $(\hbar Q)^2/2m \simeq \lvert \mu \rvert \simeq g \rho$, with $g = 4\pi \hbar^2 a/m$, with scattering length $a$ and atom mass $m$, then the energy of the entire gas after the quench is concentrated at the scale $Q \simeq k_\xi$, with healing-length momentum scale $k_\xi = \sqrt{8 \pi a \rho}$. 

Most importantly, such a strong cooling quench leads to an extreme initial condition for the subsequent dynamics. 
The post-quench distribution is strongly over-occupied at momenta $k < Q$, as compared to the final equilibrium
distribution.
This initial overpopulation of modes with energies $\sim (\hbar Q)^2 /2m$ induces inverse particle transport to lower momenta
while energy is transported to higher wavenumbers \cite{Nowak:2012gd,Orioli:2015dxa,Berges:2012us}, as indicated by the arrows in Fig.~\ref{fig:NTFP}. 
The rescaling is thus characterized by a bi-directional, in general non-local redistribution of particles and energy.
Furthermore, in contrast to the case of a weak quench leading to a scaling evolution in which, typically, weak wave turbulence is induced  \cite{Svistunov1991a,Chantesana:2018qsb.PhysRevA.99.043620}, here the inverse transport is characterized by a different, strongly non-thermal power-law form of the scaling function in the infrared (IR) region.
While the  spatio-temporal scaling provides the ``smoking gun'' for the approach of a non-thermal fixed point, in all cases examined so far, this steep power-law scaling of the momentum distribution, $n(k) \sim k^{-\zeta}$, has been observed and reflects the character of the underlying transport, see Fig.~\ref{fig:NTFP}. 
The evolution during this period is universal in the sense that it becomes mainly independent of the precise initial conditions set by the cooling quench as well as of the particular values of the physical parameters characterizing the system.

In the vicinity of the non-thermal fixed point, the momentum distribution of the Bose gas rescales self-similarly, within a certain range of momenta, according to $n(k,t)=(t/t_{\mathrm{ref}})^{\alpha}n([t/t_{\mathrm{ref}}]^{\beta}k,t_{\mathrm{ref}})$, with some reference time $t_{\mathrm{ref}}$. 
The distribution shifts to lower momenta for $\beta >0$, while transport to larger momenta occurs in the case of $\beta <0$.
A bi-directional scaling evolution is, in general, characterized by two different sets of scaling exponents.
One set describes the inverse particle transport towards  low momenta whereas the second set quantifies the transport of energy towards large momenta.

Global conservation laws -- applying within a certain, extended regime of momenta -- strongly constrain the redistribution underlying the self-similar dynamics in the vicinity of the non-thermal fixed point. Hence, they play a crucial role for the possible scaling evolution as they impose scaling relations between the scaling exponents.
For example, particle number conservation in the infrared regime of long wavelengths, in $d$ spatial dimensions, requires that $\alpha = d \beta$.

The resulting transport in momentum space can emerge from rather different underlying physical configurations and processes. 
For example, the dynamics can be driven by the conserved redistribution of quasiparticle excitations such as in weak wave turbulence \cite{Orioli:2015dxa,Chantesana:2018qsb.PhysRevA.99.043620} but also by the reconfiguration and annihilation of (topological) defects populating the system \cite{Nowak:2012gd,Karl2017b.NJP19.093014}.
The latter dynamics is often connected to the concept of superfluid turbulence, see Sec.~\ref{sec:Turbulence}.
If defects are subdominant or absent at all, e.g., in multi-component systems, the strongly occupied modes exhibiting scaling near the fixed point \cite{Orioli:2015dxa,Chantesana:2018qsb.PhysRevA.99.043620} typically reflect strong phase fluctuations not subject to an incompressibility constraint. 
They can be described by the low-energy effective theory discussed in Sec.~\ref{sec:LEEFT} \cite{Mikheev:2018adp}.
The associated scaling exponents are generically different for both types of dynamics, with and without defects
\cite{Schole:2012kt,Karl2017b.NJP19.093014,Orioli:2015dxa}.

The existence and significance of strongly non-thermal momentum power-laws, requiring a non-perturbative description reminiscent of wave turbulence was proposed by Rothkopf, Berges and collaborators in the context of reheating after early-universe inflation \cite{Berges:2008wm,  Berges:2008sr},  generalized by Scheppach, Berges, and Gasenzer to scenarios of strong matterwave turbulence \cite{Scheppach:2009wu}, and to the case of topological defects by Nowak, Sexty, Erne, Gasenzer et al.~\cite{Nowak:2010tm,Nowak:2011sk,Schole:2012kt,Nowak:2012gd,Schmidt:2012kw}, see also Refs.~\cite{Nowak:2013juc,Karl:2013mn,Karl:2013kua,Ewerz:2014tua,Karl2017b.NJP19.093014,Berges:2017ldx,Deng:2018xsk}.
Universal scaling at a non-thermal fixed point in both space and time was studied by Pi{\~n}eiro-Orioli, Boguslavski, and Berges for relativistic and non-relativistic $O(N)$-symmetric models \cite{Berges:2014xea,Orioli:2015dxa}, see also Refs.~\cite{Moore:2015adu,Berges:2015kfa}, and discussed in the context of heavy-ion collisions \cite{Berges:2013eia,Berges:2013fga,Berges:2014bba,Berges:2015ixa} as well as axionic models \cite{Berges:2017ldx}.

At this point we remark that the concept of non-thermal fixed points includes scaling dynamics which exhibits coarsening and phase-ordering kinetics \cite{Bray1994a.AdvPhys.43.357} following the creation of defects and nonlinear patterns after a quench, e.g.,  across an ordering phase transition.
We emphasize, however, that coarsening and phase-ordering kinetics in most cases are being discussed within an open-system framework, considering the system to be coupled to a heat bath. 
Moreover, most theoretical treatments of these phenomena do not take non-linear dynamics and transport into account.  

A common property of the universal evolutions is scaling behavior with evolution time as scaling parameter.
The associated scaling is reminiscent of equilibrium criticality at a continuous phase transition \cite{Goldenfeld1992a,ZinnJustin2004a,Ma2000a.ModernTheoryCritPhen}.
The system rescales in space with some power of the evolution time, which looks like zooming in or out the field of view of a microscope in real time.
To a certain extent, slowed-down dynamics and scaling in the evolution time can be seen as analogues of the universality in equilibrium critical phenomena in nonequilibrium systems
\cite{Hohenberg1977a,Bray1994a.AdvPhys.43.357,Sachdev2000a,Henkel2008a.NonEqPhaseTransitions,Taeuber2014a.CriticalDynamics}.

\begin{figure*}[t]
\begin{center}
  \includegraphics[width=0.35\textwidth]{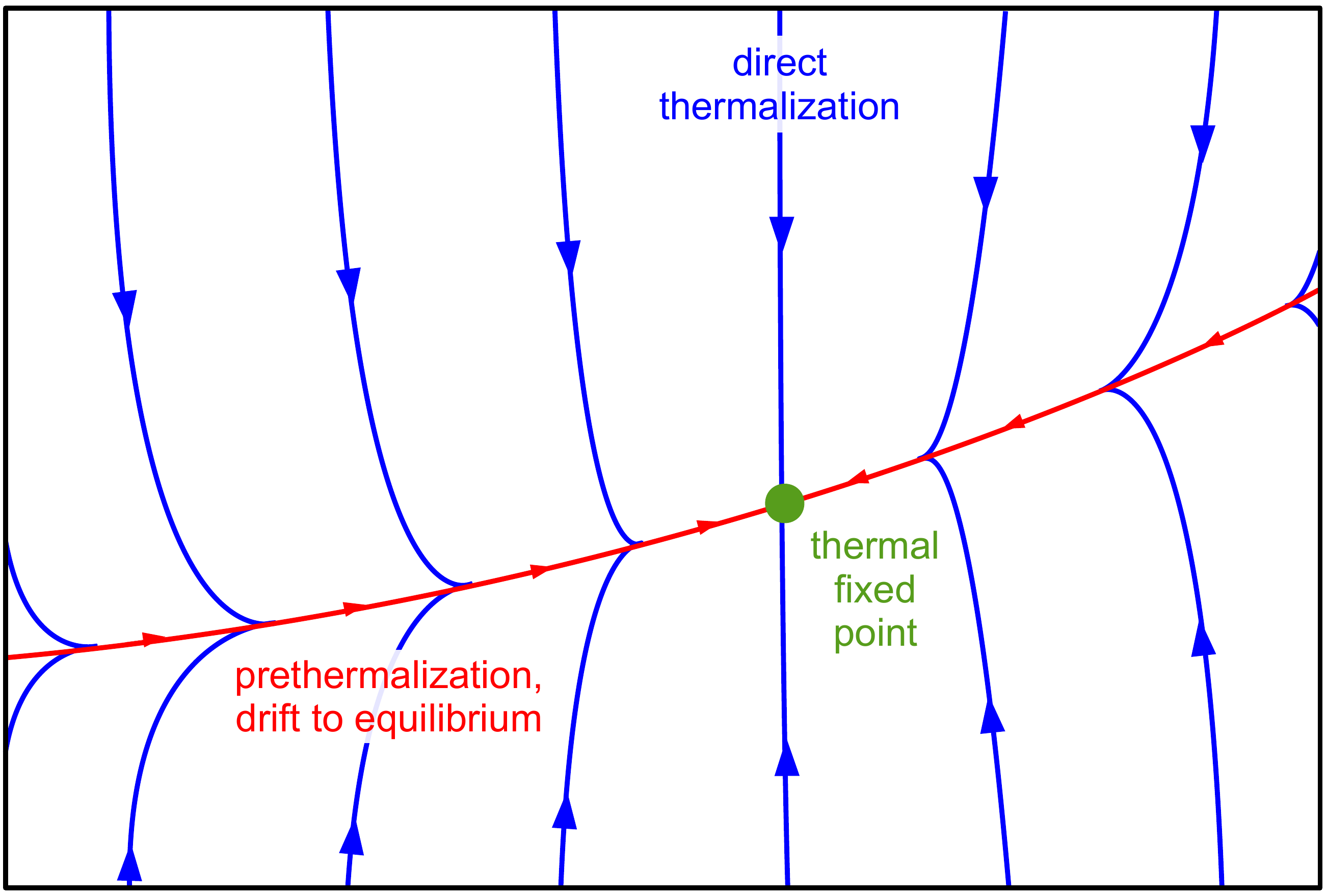}
  \hspace{0.07\textwidth}
  \includegraphics[width=0.35\textwidth]{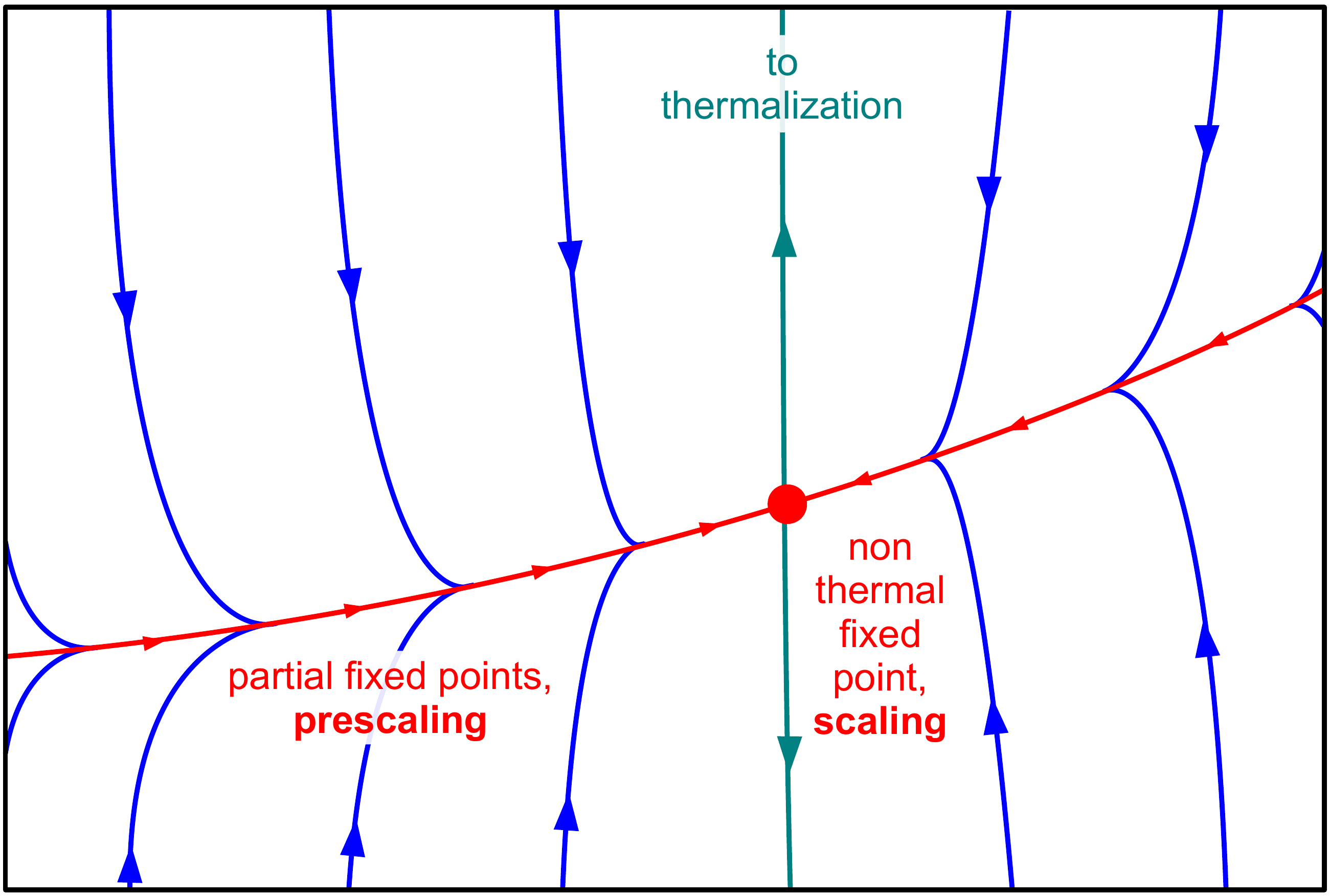}
  \caption{Schematics of prethermalization \cite{Aarts2000a.PhysRevD.63.025012,Berges:2004ce,Wetterich2012a.privcomm} (left) as compared to non-thermal fixed points (right) based on the ideas of a renormalization group flow.
The panels depict possible time evolutions in a sub-manifold of the space of many-body states.
The arbitrarily chosen axes are given by running `coupling' parameters.
These are related, e.g., to mass and coupling parameters in an effective Lagrangian, or to the various Lagrange multipliers of a generalized Gibbs ensemble (GGE).  
During the time evolution, a system can quickly approach a prethermalized state (red line). 
Due to conservation laws such a state retains memory of the initial conditions and then slowly drifts to the thermal fixed point (green). 
If, in contrast, a non-thermal fixed point is approached (right panel), prescaling, associated with partial fixed points can quickly arise, cf.~the discussion in Sec.~\ref{sec:Prescaling}.
The prescaling is associated with a symmetry being conserved during the flow while other symmetries are already broken.
Depending on the particular choice of initial condition, different, non-universal early-time evolutions (blue trajectories) occur before the system enters the universal scaling regime. 
Along the red line, correlation functions $C(k,t)$ can show approximate scaling behavior in space and time according to $C(k,t)=t^{\alpha}f(t^{\beta}k)$, with a universal scaling function $f$.
In case of prethermalization, one expects $\alpha=\beta=0$ $(\simeq0)$ such that the system gets (almost) stuck when reaching the red line. Left figure taken from Ref.~\cite{Langen:2016vdb}.}
  \label{fig:Prethermalization}
\end{center}
\end{figure*}

Prethermalization \cite{Berges:2004ce,Langen:2016vdb,Mori2018aJPhB...51k2001M} is another transient out-of-equilibrium phenomenon which is closely related to non-thermal fixed points and has originally been described using ideas from renormalization-group theory \cite{Bettencourt:1997nf,Bonini1999a,Aarts2000a.PhysRevD.63.025012,Berges:2004ce}, see Fig.~\ref{fig:Prethermalization} for an illustration. 
During the prethermalization stage, a system approaches a (partially) universal intermediate state that is, in general, still out of equilibrium with respect to asymptotically long evolution times. 
The emerging intermediate state is characterized by conservation laws relevant for the observable considered.
Universality of the state allows for a mathematical description in terms of a limited set of parameters and/or functions which solely depend on the corresponding set of symmetries obeyed in the time evolution.
Mode occupancies usually become quasi-stationary during prethermalization such that they show a trivial time evolution $\sim t^0$.
In this case one expects the associated scaling behavior of correlations to be governed by the universal scaling exponents $\alpha = \beta = 0 \, (\simeq 0)$. 
We emphasize, though, that for prethermalization, as opposed to the approach of a non-thermal fixed point, one typically does not have to take into account non-linear transport between different momentum excitations.

The characteristics of a universal intermediate configuration during the prethermalization stage are independent of the particular physical realization and of the specific initial condition.
In case of a state being only partially universal, the dynamics is dominated by the universal characteristics while non-universal properties remain.
Such non-universal properties can depend on the particular initial state of the system.

The generalized Gibbs ensemble (GGE) \cite{Jaynes1957a, Jaynes1957b,Rigol2007a.PhysRevLett.98.050405,Langen2015b.Science348.207,Vidmar2016JSMTE..06.4007V} is an example for a partially universal state occurring during the prethermalization stage.  
A GGE contains a limited set of conserved operators which are directly related to intrinsic symmetries of the Hamiltonian describing the system under consideration. 
However, the values of the Lagrange multipliers associated with the conserved quantities are non-universal as they depend on the initial values of these quantities.
The GGE is a direct generalization of thermodynamical ensembles. If only the total energy and the particle number are conserved, it reduces to the grand-canonical ensemble with Lagrange multipliers given by the temperature and the chemical potential. 
For quantum integrable systems it is believed that the GGE is the final state of relaxation \cite{Polkovnikov2011a.RevModPhys.83.863,Calabrese2016b,Eisert:2014jea,Gogolin:2016hwy,Girardeau1970a,Girardeau1969a},
while it has also been discussed in the context of wave-turbulent scaling evolution \cite{Gurarie1995a}.

\section{Kinetic theory of non-thermal fixed points}
\label{sec:KineticTheory}
In the previous section we have introduced the basic ideas of non-thermal fixed points.
For studying them theoretically we can make use of analytical as well as numerical tools.
In this section we will focus on an analytical approach to describe the scaling behavior at such fixed points.
For more technical descriptions at various levels of detail see Refs.~\cite{Berges:2004yj,Gasenzer2009a,Berges:2015kfa,Chantesana:2018qsb.PhysRevA.99.043620}.

For a general analytical treatment of non-thermal fixed points we need to be able to calculate the time evolution of a quantum many-body system out-of equilibrium.
Suitable techniques are provided by the framework of non-equilibrium quantum field theory (QFT). 
Using a path integral formulation, all information about the time-evolving quantum system is contained in the so-called Schwinger-Keldysh non-equilibrium generating functional \cite{Berges:2004yj}. 
Correlation functions, which show universal scaling at a non-thermal fixed point, can be obtained by functional differentiation of the generating functional with respect to corresponding sources.
To calculate such observables at some instant in time, the system is evolved along a Schwinger-Keldysh closed time path which reflects the nature of non-equilibrium QFT as an initial value problem.
This is in contrast to equilibrium QFT, where only asymptotic input and output states are used.
The initial configuration of the out-of equilibrium system is contained in the initial density matrix which enters the generating functional. 
In the majority of cases it is sufficient to choose the initial density matrix to be Gaussian. 
Calculating the non-equilibrium generating functional in its most general formulation is highly non-trivial.

To study the universal scaling behavior at non-thermal fixed points we focus on the evolution of two-point correlators.
From these, e.g., (quasi)particle occupation numbers in momentum space can be derived if the system is, on average, spatially translation invariant.
Taking the Schwinger-Keldysh description one derives dynamical equations for unequal-time two-point correlators, called Kadanoff-Baym equations \cite{KadanoffBaym1962a}.
These equations describe the non-equilibrium dynamics exactly but are as non-trivial to solve as the computation of the non-equilibrium generating functional entering these equations is. 
One is thus held to reduce the complexity of the problem and to obtain approximate dynamical equations that are capturing the physics relevant at a non-thermal fixed point.
It turns out that a kinetic theory approach provides such an approximation, see, e.g., Refs.~\cite{Berges:2004yj,Berges:2015kfa,Lindner:2005kv}. 

Below we will briefly outline, following Ref.~\cite{Chantesana:2018qsb.PhysRevA.99.043620}, how to proceed from the Kadanoff-Baym equations in order to derive a kinetic equation that governs the scaling behavior at a non-thermal fixed point.
As a first step, the two-point correlators are decomposed into a symmetric and asymmetric part. 
For bosons, the symmetric part, which is termed the statistical function $F$, is given by the anti-commutator of the field operators at two points in space and time, whereas the spectral function $\rho$, defined by the commutator, represents the antisymmetric part. 
To obtain a kinetic equation it is necessary to introduce a quasiparticle Ansatz for the spectral function.
In a next step a gradient expansion of the Kadanoff-Baym equations with respect to the evolution time is performed.
As universal scaling is reminiscent of a loss of memory about the initial condition, we can formally put the initial time to minus infinity and thus forget about contributions to the equations coming from the initial state. 
Taking the gradient expansion to leading order and in an equal-time limit, we obtain a Boltzmann-type kinetic equation for the time evolution of the (quasi)particle occupation number.  
This equation is termed generalized quantum Boltzmann equation (QBE). 

Using a kinetic-theory description enables us to perform a scaling analysis from which the scaling exponents associated with the non-thermal fixed point can be predicted analytically.
To outline this analytical approach we consider an $N$-component homogeneous Bose gas, with $O(N) \times U(1)$-symmetric interactions, given by the Gross-Pitaevskii Hamiltonian 
\begin{equation}
\label{eq:HamiltonianMulticomponentBoseGas}
H = \int \mathrm{d}^d x \left [- \Phi_a^\dagger \frac {\nabla^2}{2m} \Phi_a + \frac g 2 \Phi_a^\dagger \Phi_b^\dagger \Phi_b \Phi_a \right].
\end{equation}
Here, the time and space dependent fields $\Phi_a \equiv \Phi_a(\mathbf{x},t)$, $a = 1, ...,N$, satisfy Bose equal-time commutation relations, $m$ is the mass of the bosons and the contact interaction is quantified by a single coupling $g = 4 \pi \hbar^2 a/m$, defined in terms of the s-wave scattering length $a$.
Note that we use units where $\hbar = 1$ and that it is summed over the Bose fields according to Einstein's sum convention.
For  simplicity of the notation, field indices are suppressed in the following. 

Within kinetic theory, the object of interest is the occupation number distribution $n(\mathbf{k},t) = \left\langle \Phi^{\dagger}(\mathbf{k},t) \Phi(\mathbf{k},t) \right\rangle$.
As already mentioned above, the time evolution of $n(\mathbf{k},t)$ is described in terms of a generalized Quantum Boltzmann equation (QBE)
\begin{equation}
\label{eq:QBE}
\partial_t n (\mathbf{k},t) = I[n](\mathbf{k},t),
\end{equation}
where $I[n](\mathbf{k},t)$ is a scattering integral that takes the form
\begin{align}
\label{eq:scattering}
   I[n](\mathbf k,t)
  = &\int_{\mathbf p \mathbf q \mathbf r}|T_{\mathbf k \mathbf p\mathbf q \mathbf r}|^{2}\,\delta(\mathbf k+\mathbf p -\mathbf q - \mathbf r)\,
  \delta(\omega_{\mathbf k}+\omega_{\mathbf p}-\omega_{\mathbf q}-\omega_{\mathbf r})
  \nonumber\\
  &\quad \times\
  [(n_{\mathbf k}+1)(n_{\mathbf p}+1)n_{\mathbf q}n_{\mathbf r}
  -\
  n_{\mathbf k}n_{\mathbf p}(n_{\mathbf q}+1)(n_{\mathbf r}+1)].
\end{align}
Here, $T_{\mathbf{kpqr}}$ is the scattering T-matrix and the short-hand notation $\int_{\mathbf{p}} \equiv \int \mathrm{d}^d p \, (2 \pi)^{-d}$ was introduced.
The scattering integral (\ref{eq:scattering}) describes the redistribution of the occupations $n_{\mathbf{k}}$ of momentum modes $\mathbf{k}$ with eigenfrequency $\omega_{\mathbf{k}}$ due to elastic $2 \to 2$ collisions.
In presence of a Bose condensate, the occupation numbers describe quasiparticle excitations. This modifies the scattering matrix and the mode eigenfrequencies. 
Here, we consider transport entirely within the range of a fixed scaling of the dispersion $\omega_{\mathbf{k}} \sim k^z$, with dynamical scaling  exponent $z$, such that processes leading to a change in particle number are suppressed. 
We capture collective scattering effects beyond $2 \to 2$ scattering in the T-matrix (see Sec.~\ref{sec:PropScattIntAndTmatrix}).  

Two classical limits of the QBE scattering integral $I[n](\mathbf{k},t)$ exist.
The usual Boltzmann integral for classical particles is obtained in the limit of $n(\mathbf{k},t) \ll 1$.
In case of large occupation numbers $n(\mathbf{k},t) \gg 1$, termed classical-wave limit,
the scattering integral reads
\begin{align}
   I[n](\mathbf k,t)
  = &\int_{\mathbf p \mathbf q \mathbf r}|T_{\mathbf k \mathbf p\mathbf q \mathbf r}|^{2}\,\delta(\mathbf k+\mathbf p -\mathbf q - \mathbf r)\,
  \delta(\omega_{\mathbf k}+\omega_{\mathbf p}-\omega_{\mathbf q}-\omega_{\mathbf r})
  \nonumber\\
  &\quad \times\
  [(n_{\mathbf k}+ n_{\mathbf p})n_{\mathbf q}n_{\mathbf r}
  -\
  n_{\mathbf k}n_{\mathbf p}(n_{\mathbf q} + n_{\mathbf r})].
  \label{eq:KinScattIntCWL}
\end{align}
The QBE reduces to the wave-Boltzmann equation (WBE) which is subject of the following discussion as we are interested in the universal dynamics of a near-degenerate Bose gas obeying $n(\mathbf{k},t) \gg 1$ within the relevant momentum regime.

\subsection{Properties of the scattering integral and the T-matrix}
\label{sec:PropScattIntAndTmatrix}

Scaling features of the system at a non-thermal fixed point are directly encoded in the properties of the scattering integral.  
For a general treatment that governs the cases of presence and absence of a condensate density, we focus on the scaling of the   distribution of quasiparticles, in the following denoted by $n_Q(\mathbf{k})$, instead of the single-particle momentum distribution $n(\mathbf{k})$. 
Note that in case of free particles, with dispersion $\omega(k)=k^{2}/2m\sim k^{z}$, i.e.~dynamical exponent $z=2$, we obtain $n_Q \equiv n$.
For Bogoliubov sound with dispersion $\omega(k)=c_\mathrm{s}k$ and thus $z=1$, the scaling of $n_Q$ differs from the scaling of $n$ due to the $k$-dependent Bogoliubov mode functions characterizing the transformation between the particle and quasiparticle basis, $n(\mathbf{k})\simeq (g\rho_{0}/c_\mathrm{s}k)n_{Q}(\mathbf{k})$, for $k\to0$, in general $n(\mathbf{k})\sim k^{z-2+\eta}n_{Q}(\mathbf{k})$, with anomalous exponent $\eta$. Here, $c_s$ denotes the speed of sound of the Bogoliubov excitations and $\rho_0$ is the condensate density.

Using a positive, real scaling factor $s$, the self-similar evolution of the quasiparticle distribution at a non-thermal fixed point reads
\begin{equation}
\label{eq:ScalingnQ}
n_Q(\mathbf{k},t) = s^{\alpha/\beta} n_Q \left(s\mathbf{k}, s^{-1/\beta} t \right).
\end{equation}
We remark that by choosing the scaling parameter $s = (t/t_{\mathrm{ref}})^{\beta}$ we obtain the scaling form stated in the example in Sec.~\ref{sec:NTFP}.

As the scattering integral, in the classical-wave limit, is a homogeneous function of momentum and time, it obeys scaling, provided the scaling of the quasiparticle distribution in (\ref{eq:ScalingnQ}), according to
\begin{equation}
\label{eq:ScalingScattInt}
I[n_Q] (\mathbf{k}, t) =  s^{- \mu}I[n_Q] (s\mathbf{k}, s^{-1/\beta}t),
\end{equation}
with scaling exponent $\mu = 2(d+m) -z -3\alpha/\beta$. 
Here, $m$ is the scaling dimension of the modulus of the T-matrix 
\begin{equation}
\label{eq:Tscaling0}
|T(\mathbf k, \mathbf p, \mathbf q, \mathbf r; t)| = s^{-m} |T(s \mathbf k, s \mathbf p, s \mathbf q, s \mathbf r; s^{-1/\beta} t)|.
\end{equation}

At a fixed instance in time, the T-matrix can have a purely spatial momentum scaling form.
Consider a simple example of a universal quasiparticle distribution at a fixed time $t_0$, which, at least in a
limited regime of momenta, shows power-law scaling, 
\begin{equation}
\label{eq:FixedTimeScalingnQ}
n_Q(s \mathbf{k}) = s^{- \kappa} n_Q(\mathbf{k}),
\end{equation}
with fixed-time momentum scaling exponent $\kappa$.
The T-matrix is then expected to scale as 
\begin{equation}
\label{eq:Tscaling}
|T(\mathbf k, \mathbf p, \mathbf q, \mathbf r; t_0)| = s^{-m_{\kappa}} |T(s \mathbf k, s \mathbf p, s \mathbf q, s \mathbf r; t_0)|,
\end{equation} 
with $m_{\kappa}$ being, in general, different from $m$.
Note that Eq.~(\ref{eq:FixedTimeScalingnQ}) in realistic cases is regularized by an IR cutoff $k_{\Lambda}$ or, respectively, a UV cutoff $k_{\lambda}$ to ensure that the scattering integral stays finite in the limit $k \gg k_{\Lambda}$ or $k \ll k_{\lambda}$.

Generally, the scaling hypothesis for the T-matrix, Eq.~(\ref{eq:Tscaling0}), does not hold over the whole range of momenta.
In fact, scaling, with different exponents, is found within separate limited scaling regions which we discuss in the following.

\subsubsection*{Perturbative region: two-body scattering}

For the non-condensed, weakly interacting Bose gas away from unitarity the $T$-matrix is well approximated by 
\begin{equation}
  |T_{\mathbf k\mathbf p\mathbf q\mathbf r}|^{2} = (2\pi)^{4}g^{2} \,.
  \label{eq:Titogbare}
\end{equation}
As the matrix elements are momentum independent we obtain $m_\kappa = m = 0$.
It can be shown that Eq.~(\ref{eq:Titogbare}) represents the leading perturbative approximation of the full momentum-dependent many-body coupling function.

In presence of a condensate density $\rho_{0}\leq\rho$, sound wave excitations become relevant below the healing-length momentum scale  $k_{\xi}=\sqrt{2g\rho_{0}m}$. 
Within leading-order perturbative approximation, the elastic scattering of these sound waves is described by the T-matrix 
\begin{align}
  &|T_{\mathbf k\mathbf p\mathbf q\mathbf r}|^{2}
  =\ (2\pi)^{4}\frac{(mc_{s})^{4}}{kpqr} \frac{3g^{2}}{2}  \,.
  \label{eq:TitogbareQP}
\end{align}
Here, the speed of sound of the quasiparticle excitations $c_s$ is given by $mc_{s}=k_{\xi}/\sqrt{2}=\sqrt{g\rho_{0}m}$. 
For the Bogoliubov sound we obtain the scaling exponents $m_\kappa = m = -2$.

The above perturbative results are in general applicable to the UV range of momenta. 
However, scaling behavior in the far IR regime, where the momentum occupation numbers grow large, requires an approach beyond the Boltzmann, leading-order perturbative approximation as perturbative contributions to the scattering integral of order higher than $g^2$ are no longer negligible.

\begin{figure}[t]
\begin{center} 
\includegraphics[width=0.7\columnwidth]{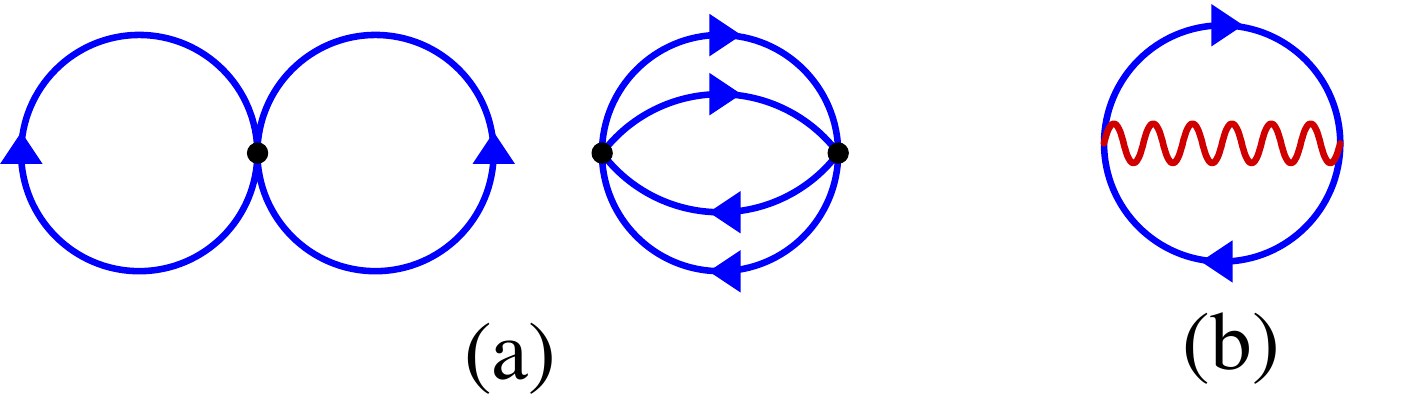}\\
\vspace{0.4cm}
\includegraphics[width=0.7\columnwidth]{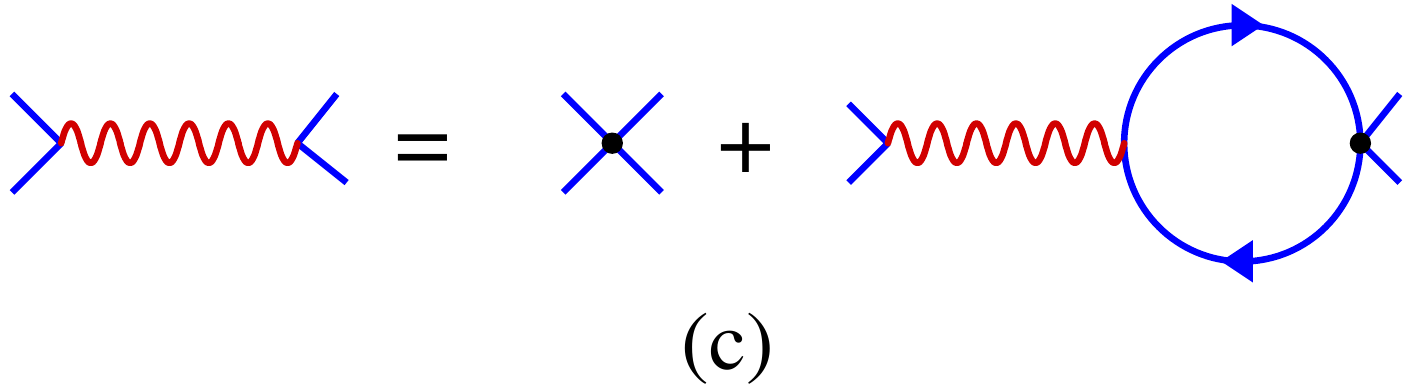}
\caption{Graphical representation of the resummation scheme.
(a) The two lowest-order diagrams of the loop expansion of the two-particle irreducible effective action (or $\Phi$ functional) which lead to the Quantum-Boltzmann equation and thus to the bare coupling $g$ within the perturbative region. 
Solid lines represent the propagator $G(x,y)$, black dots the bare vertex $\sim g\delta(x-y)$.
(b) Diagram representing the resummation approximation which replaces the diagrams in (a) within the IR regime of momenta and gives rise to the modified scaling of the $T$-matrix.
(c) The wiggly line is the effective coupling function entering the $T$-matrix, which corresponds to a sum of bubble-chain diagrams.
Figure taken from Ref.~\cite{Chantesana:2018qsb.PhysRevA.99.043620}.
\label{fig:2PI}}
\end{center}
\end{figure} 

\subsubsection*{Collective scattering: non-perturbative many-body $T$-matrix}

To do so, we use a non-perturbative s-channel loop resummation derived
within a quantum-field-theoretic approach based on the two-particle irreducible (2PI) effective action or $\Phi$-functional.
The resummation procedure is schematically depicted in Fig.~\ref{fig:2PI}.
It is equivalent to a large-$N$ approximation at next-to-leading order and enables to calculate an effective momentum-dependent coupling constant $g_{\mathrm{eff}}(k)$ which replaces the bare coupling $g$.
The effective coupling also changes the scaling exponent $m$ of the T-matrix within the IR regime of momenta.
In particular, $g_{\mathrm{eff}}(k)$ becomes suppressed in the IR to below its bare value $g$.
This ultimately leads to an even steeper rise of the (quasi)particle spectrum. 

For free particles ($z=2$) in $d=3$ dimensions we obtain
\begin{equation}
\label{eq:Titogeff}
|T_{\mathbf k \mathbf p \mathbf q \mathbf r}| = (2 \pi)^4 g_{\mathrm{eff}}^2(\varepsilon_{\mathbf{k}} - \varepsilon_{\mathbf{r}}, \mathbf{k} - \mathbf{r}),
\end{equation}
where $\varepsilon_{\mathbf{k}} - \varepsilon_{\mathbf{r}}$ and $\mathbf{k} - \mathbf{r}$ are the energy ($\varepsilon_{\mathbf{k}}=|\mathbf{k}|^{2}/2m$) and momentum transfer in a scattering process, respectively. 

\begin{figure*}[t]
    \centering
    \includegraphics[width=0.37\textwidth]{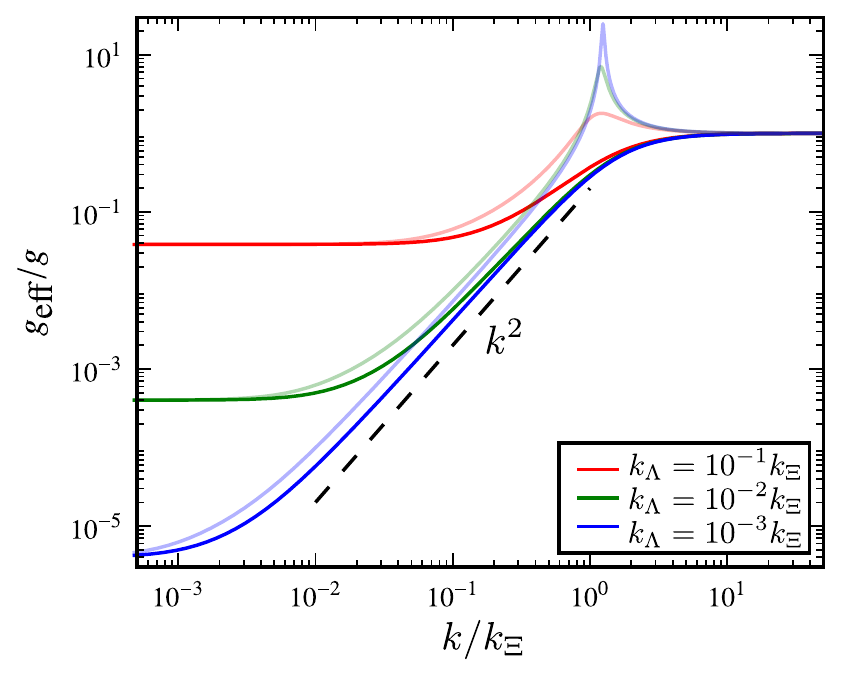}
     \hspace{0.75 cm}
     \includegraphics[width=0.37\textwidth]{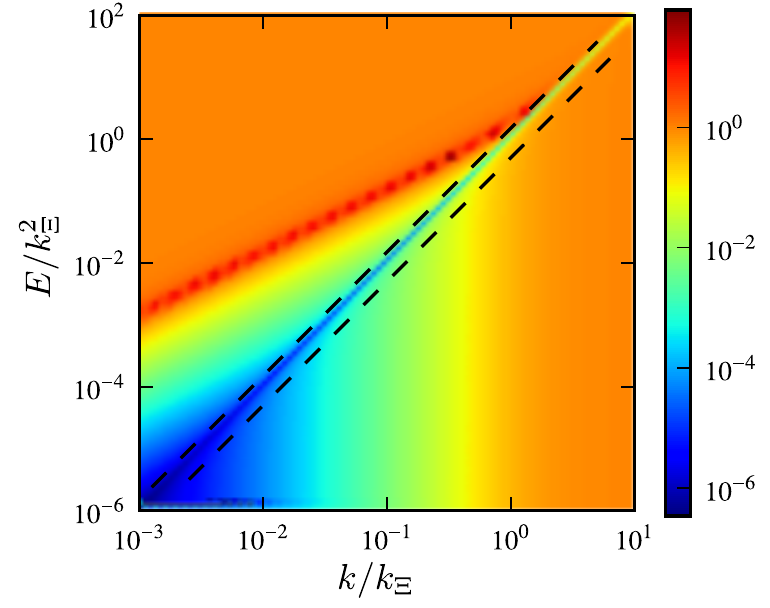}
    \caption{\textit{Left panel:} Effective coupling $g_\mathrm{eff}(k_{0},k)/g$ in $d=3$ dimensions as a function of the spatial momentum  $k=|\mathbf k|$, on a double-logarithmic scale. The graph shows cuts along $k_{0} = 0.5\varepsilon_{\mathbf k}$ (dark solid lines) and $k_{0} = 1.5\varepsilon_{\mathbf k}$ (transparent solid lines), where $\varepsilon_{\mathbf{k}}=|\mathbf{k}|^{2}/2m$. 
    Different colors correspond to different IR cutoffs $k_\Lambda$ which are set by the scaling form of the occupation number distribution entering the non-perturbative coupling function.
    All momenta are measured in units of the `healing'-length momentum scale $k_{\Xi}=(2g\rho_{\mathrm{nc}} m)^{1/2}$ of the non-condensed particle density $\rho_{\mathrm{nc}}$ under $n(k)$.
   $k_{\Xi}$ sets the scale separating the perturbative region at large momenta from the non-perturbative collective-scattering region within which the coupling assumes the form given in Eq.~(\ref{eq:geffFreeUniversal}).
    \textit{Right panel:}
    Contour plot of the effective coupling function $g_\mathrm{eff}(k_{0},k)/g$ as a function of $E=2mk_{0}$ and momentum $k=|\mathbf{k}|$. The data is depicted for $k_\Lambda = 10^{-3}k_{\Xi}$.  
	The two cuts shown in the left panel correspond to the black dashed lines. 
    The quasiparticle distribution $n_Q(k)\equiv n(k)$ was chosen to scale with $\kappa = 3.5$ such that we are in a regime where the effective coupling assumes the universal scaling form (\ref{eq:geffFreeUniversal}).
    Figures adapted from Ref.~\cite{Chantesana:2018qsb.PhysRevA.99.043620}.
    }
    \label{fig:EffCoupling}
\end{figure*}

The resulting momentum-dependent effective coupling function $g_{\mathrm{eff}}(k_{0},\mathbf k)$ along two exemplary cuts $k_{0} = 0.5\varepsilon_{\mathbf k}$ and $k_{0} = 1.5\varepsilon_{\mathbf k}$ in frequency-momentum space, for three different IR cutoffs $k_{\Lambda}$, is shown in the left panel of Fig.~\ref{fig:EffCoupling}.

At large momenta, the effective coupling is constant and agrees with the perturbative result, i.e., one finds $g_{\mathrm{eff}}=g$. However, below the characteristic momentum scale $k_{\Xi} = \sqrt{2 g \rho_\mathrm{nc} m}$, the effective coupling deviates from the bare coupling $g$. 
Within a momentum range of 
\begin{align}
   k_{\Lambda}\ll k\ll k_{\Xi}\, ,
  \label{eq:CollScattRegime}
\end{align}
the effective coupling is found to assume the universal scaling form
\begin{equation}
 \label{eq:geffFreeUniversal}
 g_{\mathrm{eff}}(k_{0},\mathbf k) 
 \simeq \frac{\left|\varepsilon_{\mathbf k}^{2}-k_{0}^{2}\right|}{2\rho_\mathrm{nc}\, \varepsilon_{\mathbf k}}\,,
 \qquad(\kappa>3)
\end{equation}
independent of both, the microscopic interaction constant $g$, and the particular value of the scaling exponent $\kappa$ of $n_Q$. 
Here, $\rho_\mathrm{nc} = \rho_\mathrm{tot} - \rho_0$ denotes the non-condensed particle density.
Below the IR cutoff, i.e., for momenta $k<k_{\Lambda}$, the effective coupling becomes constant again.
The dependence of the scaling form (\ref{eq:geffFreeUniversal}) on $E=2mk_0$ and $k$ is visualized in the right panel of Fig.~\ref{fig:EffCoupling}.

We remark that the simple universal form of the effective coupling (\ref{eq:geffFreeUniversal}) only requires a sufficiently steep power-law scaling of the quasiparticle distribution $n_Q(k) \sim k^{-\kappa}$ and an IR regularization $k_\Lambda$ ensuring finite particle number. 

Making use of the scaling properties of the effective coupling,
\begin{align}
  g_\mathrm{eff}(k_{0},\mathbf k) 
  = s^{-\gamma_{\kappa}}g_\mathrm{eff}(s^{z}k_{0},s\mathbf k)
       \,,
  \label{eq:geffscaling}
\end{align}
we obtain $\gamma_{\kappa}=0$ in the perturbative regime and $\gamma_{\kappa}=2$ in the collective-scattering regime for free particles with $z=2$.
In combination with Eq.~(\ref{eq:Titogeff}) we find the corresponding scaling exponent of the T-matrix to be $m_\kappa= 2$.
The same analysis of the effective coupling can be performed for the Bogoliubov dispersion with $z=1$. In contrast to free particles 
the scaling exponent of the T-matrix reads $m_\kappa= 0$, see Ref.~\cite{Chantesana:2018qsb.PhysRevA.99.043620} for details.

\subsection{Scaling analysis of the kinetic equation}
\label{sec:ScalingKinEq}

We are now in the position to determine the scaling properties of the Bose gas at a non-thermal fixed point. 
Here, we focus on the case of a bi-directional self-similar evolution as obtained after performing a strong cooling quench, recall the example introduced in Sec.~\ref{sec:NTFP}.
For a detailed discussion of the scaling behavior occurring after weak cooling quenches,  where only a few of the high-energy particles in the thermal tail are removed from the system, we also refer to Ref.~\cite{Chantesana:2018qsb.PhysRevA.99.043620}.

To quantify the momentum exponent $\kappa$ leading to a bi-directional scaling evolution we study the scaling of the quasiparticle distribution at a fixed evolution time as stated in Eq.~(\ref{eq:FixedTimeScalingnQ}).
As the density of quasiparticles
\begin{equation}
\label{eq:QPDens}
\rho_Q = \int \frac {\mathrm {d}^d k}{(2 \pi)^d} n_Q (\mathbf{k})
\end{equation}
and the energy density 
 \begin{equation}
 \label{eq:QPEnergy}
\epsilon_Q = \int \frac {\mathrm {d}^d k}{(2 \pi)^d} \omega_Q(\mathbf{k}) n_Q (\mathbf{k})
\end{equation}
are physical observables, they must be finite.  
We assume that the momentum distribution is isotropic, i.e., $n_Q(\mathbf{k}) \equiv n_Q(k)$ and given by a bare power-law scaling $n_Q \sim k^{-\kappa}$.
The exponent $\kappa$ then determines whether the IR or the UV
regime dominates quasiparticle and energy densities. 
For a bi-directional self-similar evolution the quasiparticle density has to dominate the IR and the energy density the UV, due to their different scaling with $k$.
Hence the scaling exponent $\kappa$ has to fulfill 
\begin{equation}
d \leq \kappa \leq d + z.
\end{equation}
Note that the quasiparticle distribution requires regularizations in the IR and the UV limits in that case as already introduced before in terms of $k_\Lambda$ and $k_\lambda$, respectively.

According to the scaling hypothesis the time evolution of the quasiparticle distribution is captured by Eq.~(\ref{eq:ScalingnQ}), with universal scaling exponents $\alpha$ and $\beta$.
Global conservation laws strongly constrain the form of the correlations in the system and the ensuing dynamics. 
Hence, they play a crucial role for the possible scaling phenomena as they imply scaling relations between the exponents $\alpha$
and $\beta$. 
Conservation of the total quasiparticle density, Eq.~(\ref{eq:QPDens}), requires
\begin{equation}
\label{eq:ConservedQPDens}
\alpha = d \beta.
\end{equation}
Analogously, if the dynamics conserves the energy density, Eq.~(\ref{eq:QPEnergy}), the relation
\begin{equation}
\label{eq:ConservedQPEnergy}
\alpha = (d + z) \beta
\end{equation}
must be fulfilled.

Obviously, the scaling relations (\ref{eq:ConservedQPDens}) and (\ref{eq:ConservedQPEnergy}) cannot both be satisfied for non-zero $\alpha$ and $\beta$ if $z \neq 0$.
This leaves us with two possibilities: Either $\alpha = \beta = 0$ or the scaling hypothesis (\ref{eq:ScalingnQ}) has to be extended to allow for different rescalings of the IR and the UV parts of the scaling function.
In the following we denote IR exponents with $\alpha$, $\beta$ and UV exponents with $\alpha^\prime$, $\beta^\prime$ respectively.
Making use of the global conservation laws as well as of the power-law scaling of the quasiparticle distribution, $n_Q \sim k^{-\kappa}$, one finds the scaling relations
\begin{equation}
\label{eq:Beta}
\alpha = d \beta,
\end{equation}
\begin{equation}
\label{eq:BetaPrime}
\beta^\prime (d+z-\kappa) = \beta (d-\kappa).
\end{equation}
This implies $\beta \beta^\prime  \leq 0$, i.e., the IR and UV
scales $k_{\Lambda}$ and $k_{\lambda}$ rescale in opposite directions. 
We remark that these relations hold in the limit of a large scaling region of momenta, i.e., for $k_{\Lambda} \ll k_{\lambda}$. 
Note that energy conservation only affects the UV shift with exponent $\beta^\prime$, Eq.~(\ref{eq:BetaPrime}), while particle conservation
gives the relation Eq.~(\ref{eq:Beta}) for the exponent $\beta$ in the IR.

With this at hand we are finally  able to derive analytical expressions for the scaling exponents based on the kinetic theory approach.
Performing the s-channel loop-resummation, the effective coupling $g_{\mathrm{eff}}$ can be expressed by the retarded one-loop self-energy $\Pi^R$, which is defined in terms of the statistical and spectral function encoding the mode occupations and, respectively, the dispersion relation as well as the density of states of the system. 
The aforementioned anomalous dimension $\eta$ appears as a scaling dimension of the spectral function.
The particle and quasiparticle distributions are obtained by frequency integrations over the statistical function. 
The resulting, most general scaling relations for the (quasi)particle distributions then read
\begin{align}
n_Q(\mathbf{k},t) &= s^{\alpha/\beta} n_Q \left (s \mathbf{k}, s^{-1/\beta}t \right), \\
n_Q(\mathbf{k},t_0) &= s^{\kappa} n_Q \left (s \mathbf{k}, t_0 \right), \\
n (\mathbf{k},t) &= s^{\alpha/\beta -\eta +2 - z} n \left (s \mathbf{k}, s^{-1/\beta} t \right), \\
n (\mathbf{k},t_0) &= s^{\kappa -\eta +2 - z} n \left (s \mathbf{k}, t_0 \right) = s^{\zeta} n \left (s \mathbf{k}, t_0 \right) \label{eq:FixedTimeScalingn}.
\end{align}
To show possible differences in the scaling behavior of the particle and quasiparticle distributions we added the relations for the particle distribution which scales as $n(\mathbf{k})\sim k^{z-2+\eta}n_{Q}(\mathbf{k})$ relative to the quasiparticle number, see beginning of Sec.~\ref{sec:PropScattIntAndTmatrix}. Note that the momentum scaling of $n(\mathbf{k})$ is characterized by the scaling exponent $\zeta$ according to Eq.~(\ref{eq:FixedTimeScalingn}). 

Zeroes of the scattering integral in the kinetic equation correspond to fixed points of the time evolution.
From a scaling analysis of the QBE one obtains the scaling relation
\begin{equation}
\label{eq:ScalingRelationMu}
\alpha = 1 - \beta \mu.
\end{equation}
Making use of the scaling of the T-matrix within the different momentum regimes as well as the global conservation laws of the system, 
one finds the scaling exponents by means of simple power counting to be 
\begin{equation}
\alpha = d/z, \quad
\beta = 1/z,
\end{equation}
\begin{equation}
\alpha^\prime = \beta^\prime (d +z), \quad
\beta^\prime = \beta(3z -4 +2 \eta)(z-4-2\eta)^{-1},
\label{eq:UVexponents}
\end{equation}
\begin{equation}
\kappa = d + (3z-4)/2 + \eta, \quad \zeta = d + z/2.
\end{equation}
On the grounds of numerical simulations in Ref.~\cite{Schachner:2016frd}, the IR scaling exponent $\beta = 1/z$ has been proposed.
Note that the exponents stated in Eq.~(\ref{eq:UVexponents}) are usually not observed as the UV region is dominated by a near-thermalized tail.
During the early-time evolution after a strong cooling quench, an exponent $\zeta \simeq  d + 1$ was seen in semi-classical simulations for $d = 3$ in Refs.~\cite{Orioli:2015dxa} and \cite{Nowak:2012gd}, for $d = 2$ in Ref.~\cite{Nowak:2011sk}, and for $d = 1$ in Ref.~\cite{Schmidt:2012kw}. 
Numerically evaluating the kinetic equation in $d = 3$ dimensions also resulted in $\kappa \simeq 4$, see Ref.~\cite{Walz:2017ffj}.
For a single-component Bose gas in $d=3$ dimensions, the IR scaling exponents have recently been numerically determined to
be $\alpha = 1.66(12)$, $\beta = 0.55(3)$, in agreement with the analytically predicted values \cite{Orioli:2015dxa}.

For the Bose gas, the above stated exponents are expected to be valid in $d = 3$ dimensions as well as in $d = 2$.
The one-dimensional case is rather different due to kinematic constraints on elastic $2 \to 2$ scattering from energy and particle-number conservation.
We finally remark that the development of non-linear and topological excitations in combination with strong phase coherence is likely to modify the results presented, potentially through an appropriate modification of the scaling exponents $z$ and $\eta$.

\section{Low-energy effective field theory}
\label{sec:LEEFT}
In the previous section, collective phenomena that modify the properties of the scattering matrix were taken into account by means of a non-perturbative coupling resummation scheme. 
Alternatively, one can think of the idea to reformulate the theory in terms of new degrees of freedom in the first place, such that the resulting description becomes more easy to treat in non-perturbative regions. 
Since the non-perturbative behavior appears at low momentum scales, it is suggestive to use a low-energy effective field theory (LEEFT) approach~\cite{Georgi:1994qn,Pich:1998xt}. 
This typically implies a choice of suitable degrees of freedom describing the physics occurring below a chosen energy scale. 
Classical examples of low-energy effective field theories include the Fermi theory of $\beta$-decay~\cite{Fermi2008}, 
the BCS theory of superconductivity~\cite{Bardeen1957a,Bardeen1957b} and the XY-model of superfluidity~\cite{Wen2004a.QFT}. Even approaches to quantum gravity can be made using low-energy effective field theories \cite{Donoghue:1994dn}. 
In the following, we will outline the (Wilsonian) LEEFT approach to the description of non-thermal fixed points in a multicomponent Bose gas \cite{Mikheev:2018adp}. The  ideas are based on the treatment of the aforementioned XY-model.  

The key observation is, that the $N$-component Gross-Pitaevskii model, see Eq.~(\ref{eq:HamiltonianMulticomponentBoseGas}), offers a natural separation of scales. 
An analysis of the classical equations of motion, obtained within a density-phase representation of the field, $\Phi_{a}=\sqrt{\rho_{a}}\exp\{\theta_{a}\}$, shows that, at low momenta, density fluctuations $\delta \rho_a=\rho_{a}-\rho_{a}^{(0)}$ around a mean density $\rho_{a}^{(0)}$ are suppressed by a factor of $\sim |\mathbf{k}|/k_{\Xi}$ compared to phase fluctuations $\theta_a$ (around a constant background phase). 
Here, $k_{\Xi} = [2 m \rho^{(0)} g]^{1/2}$ is the healing-length momentum scale associated with the total density $\rho^{(0)}=\sum_{a}\rho^{(0)}_{a}$. 
Hence, density fluctuations can be integrated out to obtain the low-energy effective action $S_{\mathrm{eff}}$  of the system. 

Furthermore, the model provides two types of eigenmodes: $N-1$ Goldstone excitations with a free-particle-like dispersion $\omega_1(\mathbf{k}) = ... =\omega_{N-1}(\mathbf{k}) = \mathbf{k}^2/2m$, which correspond to relative phases between different components, and a single Bogoliubov quasiparticle mode with $\omega_N(\mathbf{k}) = \left[{\mathbf{k}^2}/{2m} \left({\mathbf{k}^2}/{2m} + 2 g \rho^{(0)} \right)\right]^{1/2}$ related to the total phase. 
This suggests that the physics below the scale $k_{\Xi}$ is well-described by the dynamics of phonon-like quasiparticles, although two sorts of quasiparticles are present. 

The low-energy effective action corresponding to these quasiparticles contains interaction terms with momentum-dependent couplings indicating the fact that the resulting theory is non-local in nature, as is expected for a LEEFT~\cite{Mikheev:2018adp}. Moreover, taking the large-$N$ limit, this action becomes diagonal in component space up to $\mathcal{O}(1/N)$ corrections and thus breaks up into $N$ independent replicas. 
This means that the phases $\theta_a$ of the different components decouple in the limit of large $N$. 
Taking the limit $N \to \infty$, the Bogoliubov mode is no longer present suggesting that relative phases are dominating the dynamics of the system.
The $N \to \infty$ effective action in momentum space is found to be \cite{Mikheev:2018adp}
\begin{align}
  &S_{\mathrm{eff}}[\theta] 
  = \int_{\mathbf{k}, \mathbf{k}', \mathcal{C}} \frac{1}{2} \,\theta_a (\mathbf{k}, t) i D_{{ab}}^{-1}(\mathbf{k}, t; \mathbf{k}', t') \theta_b(\mathbf{k}', t')
  \nonumber\\
  &- \int_{\lbrace \mathbf{k}_i \rbrace, \mathcal{C}} 
  \frac{\mathbf{k}_1 \cdot \mathbf{k}_2}{2mN\,g_{\mathrm{1/N}} (\mathbf{k}_3)} \, 
  \theta_a (\mathbf{k}_1, t)\, \theta_a (\mathbf{k}_2, t) \partial_t \theta_a (\mathbf{k}_3, t) \,\delta \Big(\sum_{i=1}^{3} \mathbf{k}_i\Big) 
  \nonumber\\
  &+ \int_{\lbrace \mathbf{k}_i \rbrace, \mathcal{C}} 
  \frac{(\mathbf{k}_1 \cdot \mathbf{k}_2) \, (\mathbf{k}_3 \cdot \mathbf{k}_4)}{8m^{2}N\,g_{\mathrm{1/N}} (\mathbf{k}_1 - \mathbf{k}_2)} \, \theta_a(\mathbf{k}_1,t) \cdots \theta_a(\mathbf{k}_4,t)\, \delta \Big(\sum_{i=1}^{4} \mathbf{k}_i\Big)\,.
  \label{eq:Seff4}
\end{align} 
Here, $\mathcal{C}$ indicates the integration over a Schwinger-Keldysh contour and $D_{{ab}}$ is a free inverse propagator 
\begin{align}
i D_{{ab}}^{-1}(\mathbf{k}, t; \mathbf{k}', t') 
= \frac{(2 \pi)^d \, \delta (\mathbf{k} + \mathbf{k}')}{Ng_{\mathrm{1/N}} (\mathbf{k})} \, \delta_{{ab}} \delta_{\mathcal{C}} (t - t') \left(-\partial_t^2 - (\mathbf{k}^2/2m)^2 \right)\,.
\end{align}
In the above expression, the momentum-depending coupling $g_{\mathrm{1/N}} (\mathbf{k}) = g \mathbf{k}^2/2 k_{\Xi}^2 \equiv g_{\mathrm{G}}(\mathbf{k})/N$ was introduced, which remarkably coincides with the universal coupling obtained in the non-perturbative resummation within the 2PI formalism \cite{Chantesana:2018qsb.PhysRevA.99.043620}, see Sec.~\ref{sec:KineticTheory}. The index $G$ of the coupling refers to the relevant Goldstone excitations in the  large-N limit.

\subsection{Spatio-temporal scaling}
\label{sec:LEEFTSpatioTempScaling}

To analyze the scaling behavior at a non-thermal fixed point we proceed as in Sec.~\ref{sec:KineticTheory} by evaluating the QBE in Eq.~(\ref{eq:QBE}).
Instead of the quasiparticle distribution $n_Q$ we consider the distribution of phase-excitation quasiparticles $f_a(\mathbf{k},t) = \langle \theta_a (\mathbf{k},t) \theta_a (-\mathbf{k},t) \rangle$. 
We again drop the indices in the following to ease the notation.
The scattering integral has two contributions arising from 3- and 4-wave interactions in the effective action action (\ref{eq:Seff4}),
\begin{align}
  I[f](\mathbf{k},t)=&\ I_{3}(\mathbf{k},t)+I_{4}(\mathbf{k},t)\,.
\end{align}
The form of the 3- and 4-point scattering integrals can be inferred from the effective action to be
\begin{align}
  I_3 (\mathbf{k},t) 
  \sim& \int_{\mathbf{p},\mathbf{q}} |T_3(\mathbf{k},\mathbf{p},\mathbf{q})|^2 \, \delta (\mathbf{k} + \mathbf{p} - \mathbf{q}) \, \delta (\omega_{\mathbf{k}} + \omega_{\mathbf{p}} - \omega_{\mathbf{q}}) 
  \nonumber\\ 
  &\times \Big[ (f_{\mathbf{k}} + 1) (f_{\mathbf{p}} + 1) f_{\mathbf{q}}  -  f_{\mathbf{k}} f_{\mathbf{p}} (f_{\mathbf{q}} + 1) \Big]\,,
  \label{eq:I_3}
  \\
  \label{eq:I_4}
  I_4 (\mathbf{k},t) 
  \sim& \int_{\mathbf{p},\mathbf{q},\mathbf{r}} |T_4(\mathbf{k},\mathbf{p},\mathbf{q},\mathbf{r})|^2 \, \delta (\mathbf{k} + \mathbf{p} - \mathbf{q} - \mathbf{r}) \, \delta (\omega_{\mathbf{k}} + \omega_{\mathbf{p}} - \omega_{\mathbf{q}} - \omega_{\mathbf{r}}) 
  \nonumber\\ 
  &\times \Big[ (f_{\mathbf{k}} + 1) (f_{\mathbf{p}} + 1) f_{\mathbf{q}} f_{\mathbf{r}}  -  f_{\mathbf{k}} f_{\mathbf{p}} (f_{\mathbf{q}} + 1) (f_{\mathbf{r}} + 1) \Big]\,,
\end{align}
where the corresponding $T$-matrices are given by
\begin{align}
\label{eq:T_3}
   |T_3(\mathbf{k},\mathbf{p},\mathbf{q})|^2 &= |\gamma(\mathbf{k},\mathbf{p},\mathbf{q})|^2 
   \frac{g_{\mathrm{G}} (\mathbf{k})\, g_{\mathrm{G}} (\mathbf{p})\, g_{\mathrm{G}} (\mathbf{q})}
   {8 \,\omega (\mathbf{k}) \, \omega (\mathbf{p}) \, \omega (\mathbf{q})}\,,\\
\label{eq:T_4}
  |T_4(\mathbf{k},\mathbf{p},\mathbf{q},\mathbf{r})|^2 &= |\lambda(\mathbf{k},\mathbf{p},\mathbf{q},\mathbf{r})|^2   
  \frac{g_{\mathrm{G}} (\mathbf{k}) \cdots g_{\mathrm{G}} (\mathbf{r})}
  {2 \omega (\mathbf{k}) \cdots 2 \omega (\mathbf{r})}\,,
\end{align}
with interaction couplings 
\begin{align}
  \gamma(\mathbf{k},\mathbf{p},\mathbf{q}) 
  &= \frac{(\mathbf{k} \cdot \mathbf{p})\, \omega(\mathbf{q})}{m\,g_{\mathrm{G}} (\mathbf{q})} + \text{perm}^{\text{s}}\,,
  \label{eq:gamma3}
  \\
  \lambda(\mathbf{k},\mathbf{p},\mathbf{q},\mathbf{r}) 
  &= \frac{(\mathbf{k} \cdot \mathbf{p}) (\mathbf{q} \cdot \mathbf{r})}{2m^{2}\,g_{\mathrm{G}} (\mathbf{k} - \mathbf{p})} 
  + \text{perm}^{\text{s}}\,.
  \label{eq:lambda4}
\end{align}
Here, `perm$^{\mathrm{s}}$' denote permutations of the sets of momentum arguments.
The scattering integrals scale, analogously to Eq.~(\ref{eq:ScalingScattInt}), with exponents
\begin{align}
\mu_3 =& \,d + 4 - 2z + \gamma - 2 \alpha/\beta,\\
\mu_4 =& \,2d + 8 - 5z + 2\gamma - 3 \alpha/\beta,
\end{align}
where $\gamma = 2(z - 1)$ is the scaling exponent of the effective coupling $g_{\mathrm{eff}}(\mathbf{k}) = s^{-\gamma} g_{\mathrm{eff}} (s \mathbf{k})$. We remark that the subscript of the coupling is chosen as a general notation covering both cases of $z=2$ as well as $z=1$.

Using the scaling relation in Eq.~(\ref{eq:ScalingRelationMu}) one can, in principle, derive a closed system of equations allowing to determine  the scaling exponents $\alpha$ and $\beta$. 
However, since, for different values of the dimensionality $d$ and the momentum scale of interest, one term in the scattering integral can dominate over the other one, it is more reasonable to analyze them independently. 
To close the system of equations, an additional relation is then required, which can be provided by either quasiparticle number conservation, Eq.~(\ref{eq:ConservedQPDens}), or energy conservation, Eq.~(\ref{eq:ConservedQPEnergy}), within the scaling regime. 
Taking these constraints into account we obtain
\begin{align}
l = 3: \qquad \beta &= \frac{1}{4 - 2z + \gamma}, \quad \beta' = \frac{1}{4 - 3z + \gamma},\\
l = 4: \qquad \beta &= \frac{1}{8 - 5z + 2\gamma}, \quad \beta' = \frac{1}{8 - 7z + 2\gamma}.
\end{align}
In the large-N limit ($z = 2$, $\gamma = 2$), the resulting scaling exponents read 
\begin{align}
\beta = 1/2, \quad \alpha = d/2
\end{align}
for both, 3- and 4-point vertices, and
\begin{align}
\beta' = -1/2, \quad \alpha' = -(d+z)/2
\end{align}
for the 4-point vertex, while, at the same time, for the 3-point vertex, no valid solution exists. We point out that the above exponents are equivalent to the respective exponents derived in the large-$N$ resummed kinetic theory for the fundamental Bose fields, for the case of a dynamical exponent $z = 2$, and a vanishing anomalous dimension $\eta = 0$, cf.~Sec.~\ref{sec:ScalingKinEq}.

One can ask whether both 3- and 4-wave interactions are equally relevant. To answer this question, a comparison of the spatio-temporal scaling properties of the scattering integrals, for a given fixed-point solution $f(\mathbf{k}, t)$, is required. Focusing on the conserved IR transport of quasiparticles, for which $\alpha = d \beta$, we obtain
\begin{align}
\label{eq:mu3QP}
-\mu_3 &= d - 2,\\
\label{eq:mu4QP}
-\mu_4 &= d - 4 + z.
\end{align}
In the large-$N$ limit, for which $z=2$, one finds $\mu_3 = \mu_4$. Hence, the relative importance of the scattering integrals $I_3$ and $I_4$ should remain throughout the evolution of the system.

\subsection{Scaling solution}
In the following we briefly discuss the purely spatial momentum scaling. 
The scaling of the QBE at a fixed evolution time $t = t_0$ implies $\kappa = - \mu_{\kappa,l}$, where $\mu_{\kappa,l}$ is the spatial scaling exponent of the corresponding scattering integral, $I_{l}(\mathbf{k}, t_{0}) =  s^{- \mu_{\kappa,l}}I_{l}(s\mathbf{k}, t_{0})$.
Power-counting of the scattering integrals, together with the above stated scaling relation, gives 
\begin{align}
\label{eq:mu3kl}
\kappa_3 &= -\mu_{\kappa,3}=4 + d + \gamma - 2z,\\
\label{eq:mu4kl}
\kappa_4 &= -\mu_{\kappa,4}=4 + d + \gamma - 5z/2.
\end{align}
For a given $\kappa_l$, and assuming the large-$N$ limit ($z = 2$ and $\gamma = 2$), one finds that
\begin{align}
\mu_{\kappa,3} - \mu_{\kappa,4} = \kappa_l - d \geq 1.
\end{align}
Hence, the 4-wave scattering integral is expected to dominate at small momenta, $k \to 0$.
This implies that, at the non-thermal fixed point, the quasiparticle distribution $f(\mathbf{k}, t) \sim k^{-\kappa}$ is characterized by the momentum scaling exponent $\kappa = \kappa_4 = d + 1$. 
The result appears to contradict the previous analysis of the spatio-temporal scaling, which, in the large-$N$ limit, showed equal importance of $I_3$ and $I_4$. 
We emphasize, however, that the scaling exponents $\alpha$ and $\beta$ corresponding to the spatio-temporal scaling properties are obtained from relations which are independent of the precise form of $f(\mathbf{k}, t)$ but only require the scaling relation $f(\mathbf{k},t) = (t/t_{\mathrm{ref}})^{\alpha} f([t/t_{\mathrm{ref}}]^{\beta} \mathbf{k})$. 
Hence, the questions which vertex is responsible for the shape of the scaling function and which of the vertices dominates the transport can be answered independently of each other. 
See Ref.~\cite{Mikheev:2018adp} for further discussion.

\subsection{The case of a single-component gas, $N=1$}
While originally being derived for the limiting case of $N \to \infty$, one can also apply the LEEFT to a single-component ($N=1$) Bose gas. 
In this case, the theory describes, in the IR limit, the scattering of modes with linear Bogoliubov dispersion ($z = 1$ and $\gamma = 0$). The scaling exponent $\beta$ then reads
\begin{align}
l = 3: \qquad \beta &= 1/2,\\
l = 4: \qquad \beta &= 1/3.
\end{align}
In addition, the relation $-\mu_3 = d - 2 > -\mu_4 = d - 3$ is found. 
Hence, as the evolution time increases, the scattering integral $I_3$, for Bogoliubov-like quasiparticles, starts to win against $I_4$ such that the value of the scaling exponent $\beta$ is predicted to be $\beta = 1/2$. 
We can conclude then that transport of Bogoliubov quasiparticles towards the IR, dominated by $1 \to 2$ and $2 \to 1$ interaction processes, is described by the scaling exponents $\alpha = d/2$ and $\beta = 1/2$.

Following the same procedure as for $z=2$ one can determine the fixed-time momentum scaling exponent $\kappa$. 
The analysis reveals that 4-wave interactions dominate such that $\kappa = \kappa_4 = d + 3/2$. 
However, keeping in mind that $I_3$ becomes more relevant as time increases, one rather expects, in the $N=1$ case, the 3-wave interaction to dominate the purely spatial scaling fixed-point equation as well. Under these circumstances, the theory rather predicts the exponent 
\begin{equation}
\kappa = \kappa_3 = d + 2 \qquad\mbox{($z=1$)}
\end{equation}
to represent the momentum scaling in the long-time scaling limit.

\subsection{Relation to predictions of the non-perturbative kinetic theory}
As already pointed out, the scaling exponents derived within the LEEFT remarkably coincide with earlier findings from non-perturbatively resummed kinetic theory \cite{Chantesana:2018qsb.PhysRevA.99.043620}.
In principle, however, there is a priori no reason for them to coincide since they correspond to different degrees of freedom. 
One needs therefore a translation between the quasiparticle distribution $f_a(\mathbf{k}, t)$ characterizing the phase-angle excitations and the scaling of the particle number distribution
$n_a(\mathbf{k}, t)$ encoded in the fundamental Bose field. 
Under the assumption that the non-thermal fixed point is Gaussian with respect to quasiparticle excitations in the phase degree of freedom, it can be shown that the scaling properties derived in the large-$N$ limit,
defined by the exponents $z = 2$, $\alpha = d/2$, $\beta = 1/2$, and $\kappa = d + 1$, are consistent with the scaling properties derived within the non-perturbative
approach, for the case of a dynamical exponent $z = 2$, and a vanishing anomalous dimension $\eta = 0$.

While the idea that a non-thermal fixed point is Gaussian may sound odd in the first place, a simple scaling analysis shows that this can indeed be the case.
The Schwinger-Keldysh action $S(t)=\int_{\mathcal{C},t_\mathrm{ref}}^{t}dt'L(t')$ integrated from a reference time $t_\mathrm{ref}$ to the present  time $t$, according to the scaling hypothesis, should have a form
\begin{align}
S(t)=S(s^{1/\beta}t_\mathrm{ref})= s^{-d_{S}}S(t_\mathrm{ref}),
\end{align}
with canonical scaling dimension $d_{S}=[S]$ and scale parameter $s=(t/t_\mathrm{ref})^{\beta}$.

Using the dynamical canonical scaling dimension of the phase-angle field $\theta_{a}$ at the non-thermal fixed point, $[\theta_{a}] = -\alpha/2\beta=-d/2$, we obtain the canonical scaling of the quadratic, cubic and quartic parts of the effective action,
\begin{align}
  [S^{(2)}] &= 2z-\gamma-1/\beta &=&\ 0\,,
  \label{eq:canScDimS2}
  \\
  2[S^{(3)}] &= d+4+2z-2\gamma-2/\beta&=&\ d+4-2z\,,
  \label{eq:canScDimS3}
  \\
  [S^{(4)}] &= d+4-\gamma-1/\beta&=&\ d+4-2z\,,
  \label{eq:canScDimS4}
\end{align}
where the 3-vertex was taken twice as it occurs in even multiples in any diagram contributing to the proper self-energy, and inserted, in the respective second equations, $\gamma=2(z-1)$ and $\beta=1/2$.

If the non-thermal fixed point is Gaussian in the IR scaling limit, the conditions
\begin{align}
  2[S^{(3)}]>[S^{(2)}]\,,\qquad
  [S^{(4)}]>[S^{(2)}]
  \label{eq:GaussianFPcond}
\end{align}
need to be fulfilled, which is the case in any dimension $d>0$, as $d+4-2z>0$ for $z<(d+4)/2$.

The Gaussianity of the fixed point is also supported by the fact that the integrals $I_{l}[f]$ scale to zero in the infinite-time limit, $I_{l}[f](\mathbf{k},t)=(t/t_{\mathrm{ref}})^{(\mu_{\kappa,l}-\mu_{l})\beta} I_{l}[f](\mathbf{k},t_{\mathrm{ref}})$.
This can be inferred from their scaling exponents $\mu_{l}$ defined in Eqs.~(\ref{eq:mu3QP})--(\ref{eq:mu4kl}): 
For $z=2$, one obtains  $\mu_{3}=\mu_{4}=2-d$, $-\mu_{\kappa,3}=d+2$, $-\mu_{\kappa,4}=d+1$, such that $(\mu_{\kappa,l}-\mu_{l})\beta\leq-3/2$, independent of $d$.

We remark that to improve upon the above performed analysis, time-dependent correlation functions  need to be analyzed, e.g., within a functional renormalization-group approach \cite{Pawlowski:2005xe,Gasenzer:2008zz,Berges:2008sr}. Furthermore, we emphasize that Gaussianity of the non-thermal fixed point here refers to phase quasiparticles only, while in terms of the fundamental fields the fixed point can easily appear to be non-Gaussian.

\section{Wave-turbulent transport}
\label{sec:Turbulence}
In the previous sections we have discussed self-similar scaling dynamics at a non-thermal fixed point.
Such dynamics is characterized by bi-directional, non-local transport of particles or energy leaving a global quantity such as particle or energy density invariant in time.
This is reminiscent of the transport and scaling characterizing wave-turbulent cascades.
In these cascades, analogously to fluid turbulence, universal scaling is expected in a certain interval of momenta, termed the inertial range.
Within the inertial range of a wave-turbulent cascade, transport occurs \emph{locally}, from momentum shell to momentum shell, leaving the transported quantity within such a momentum shell constant in time.
This transport process can be described by a continuity equation in momentum space.

In a dilute Bose gas, quantities other than the kinetic energy can be locally conserved in their transport through momentum space.
In contrast to fluid turbulence, this is due to the compressibility of the gas which allows a variety of wave turbulence phenomena to arise \cite{Zakharov1992a, Nazarenko2011a}.
Taking into account particle number and energy as alternative possible conserved quantities, the respective continuity equations characterizing the local conservation laws are written as  \cite{Zakharov1992a}
\begin{eqnarray}
\label{eq:RadParticleDensity}
\partial_t N(k,t) &=& - \partial_k Q(k),
\\
\label{eq:RadEnergy}
\partial_t E(k,t) &=& - \partial_k P(k).
\end{eqnarray}
These continuity equations impose relations between the temporal change of a density and the momentum divergence of a current. 
In particular, Eq.~(\ref{eq:RadParticleDensity}) describes the time evolution of the radial particle number $N(k) = (2k)^{d-1} \pi n(k)$ driven by the radial particle current 
$Q(k) = (2k)^{d-1} \pi Q_k (k)$. 
The evolution of the radial energy $E(k) = (2k)^{d-1} \pi \omega(k) n(k)$, Eq.~(\ref{eq:RadEnergy}), where $\omega(k)$ is the dispersion of mode excitations with momentum $k$, is determined by the  radial energy current $P(k) = (2k)^{d-1} \pi P_k (k)$. 

Scaling solutions of the continuity equations are generally studied within a wave-Boltzmann kinetic approach.
Wave-turbulent scaling behavior is obtained if $Q(k)$ and $P(k)$ are independent of momentum $k$ which corresponds to a stationary distribution of particles or energy, respectively.

Weak wave turbulence is the mathematically best-controlled case.
It is found within the perturbative region, i.e., in the UV regime of momenta, where the mode occupancies are sufficiently small such that the usual wave-Boltzmann equation is valid, meaning that the scattering T-matrix is solely given by the bare coupling $g$, see also Sec.~\ref{sec:PropScattIntAndTmatrix}.
One finds a  direct relation between the continuity equations and the QBE in Eq.~(\ref{eq:QBE}) \cite{Zakharov1992a}.
The theory of weak wave turbulence therefore rests on the analysis of stationary solutions of the QBE.
By means of power counting the UV scaling exponents characterizing the weak wave turbulence are found to be \cite{Zakharov1992a,Chantesana:2018qsb.PhysRevA.99.043620}
\begin{equation}
\label{eq:ZetaUV}
\zeta^{\mathrm{UV}}_Q = d - 2/3, \qquad \zeta^{\mathrm{UV}}_P = d.
\end{equation}

In general, the energy flux corresponds to a direct cascade to larger $k$, whereas the particle flux constitutes an inverse cascade.
The character of the fluxes is entirely determined by the properties of the physical system.
We emphasize that also the self-similar evolution at a non-thermal fixed point can be described by transport equations (\ref{eq:RadParticleDensity}), (\ref{eq:RadEnergy}), however, in general with a non-local flux such that the quantity being transported does not remain constant within a given momentum shell.

Given a positive scaling exponent $\zeta$, momentum occupation numbers  $n(k) \sim k^{-\zeta}$ eventually grow
large in the deep IR. 
In that limit, the flux enters the collective scattering region where the effective many-body T-matrix is given by a non-perturbative coupling $g_{\mathrm{eff}} (k)$ and weak wave turbulence ceases to be valid,
see Ref.~\cite{Chantesana:2018qsb.PhysRevA.99.043620} for details.

The concept of wave turbulence, as well as the description of spatio-temporal scaling near a non-thermal fixed point, are based on a wave-Boltzmann kinetic approach and thus on a quasiparticle Ansatz for the excitations in the system.
While such an approach has been developed also for the IR regime of strong occupancies, cf.~Secs.~\ref{sec:KineticTheory} and \ref{sec:LEEFT}, it does not account for the effects of (quasi) topological excitations.
These excitations are observed in many different systems such as single- and multicomponent dilute Bose gases in different dimensionalities,  cf.~the corresponding numerical results discussed in Sec.~\ref{sec:Numerics}.
Analytical predictions for the respective scaling exponents and functions derived from first principles are a subject of current research in the field.

\section{Topological defects vs.~the role of fluctuations}
\label{sec:Numerics}
While we have focussed, so far, on analytical treatments of non-thermal fixed points, we present, in the following, numerical simulations of dilute Bose gases, corroborating analytical predictions as well as showing phenomena beyond analytics.
As pointed out in Sec.~\ref{sec:NTFP}, universal scaling at a non-thermal fixed point can be driven by either (quasi) topological defects populating the system or by strong fluctuations of the phase, if defects are subdominant or absent.
In the following we first present results obtained in vortex dominated single-component Bose gases. 
We will then show universal scaling at a non-thermal fixed point caused by relative-phase fluctuations in an $(N=3)$-component Bose gas.

As the scaling behavior occurs in a regime of strongly occupied modes, the time evolution can be computed by means of semi-classical simulations. 
Using the so-called truncated Wigner approximation  \cite{Blakie2008a,Polkovnikov2010a}, we follow the evolution, starting from a noisy initial configuration, by evaluating many trajectories according to the classical equations of motion.
In the simplest case of a single-component Bose gas, the equation of motion is the Gross-Pitaevskii equation
\begin{equation}
\label{eq:GPE}
i \partial_t \Phi(\mathbf{x},t) = \left [ - \frac {\nabla^2}{2m} + g \lvert \Phi(\mathbf{x},t) \rvert^2 \right] \Phi(\mathbf{x},t).
\end{equation}

Eq.~(\ref{eq:GPE}) can be mapped to a continuity equation for the density $\rho=|\Phi|^{2}$ and an Euler-type hydrodynamic equation for the superfluid velocity $\mathbf{v}=\nabla\arg(\Phi)/m$ such that the dynamics of the Bose gas can be interpreted in terms of superfluid flow.
For positive $g$, possible solutions of this equation include topological  configurations such as (dark) solitons and vortices \cite{Pismen1999a,Pitaevskii2003a}.
Solitons are quasi-topological defects which in general travel with a fixed velocity but are non-dispersive, i.e., stationary in shape and stable in $d=1$ dimension. 
Vortices are topologically stable solutions in $d > 1$ dimensions which form the superfluid analogies of eddy flows in classical fluids.
In $d = 3$ dimensions, point vortices extend to vortex lines or loops around which the fluid rotates.

 \begin{figure}[t]
 \center
  \includegraphics[width=0.75\columnwidth]{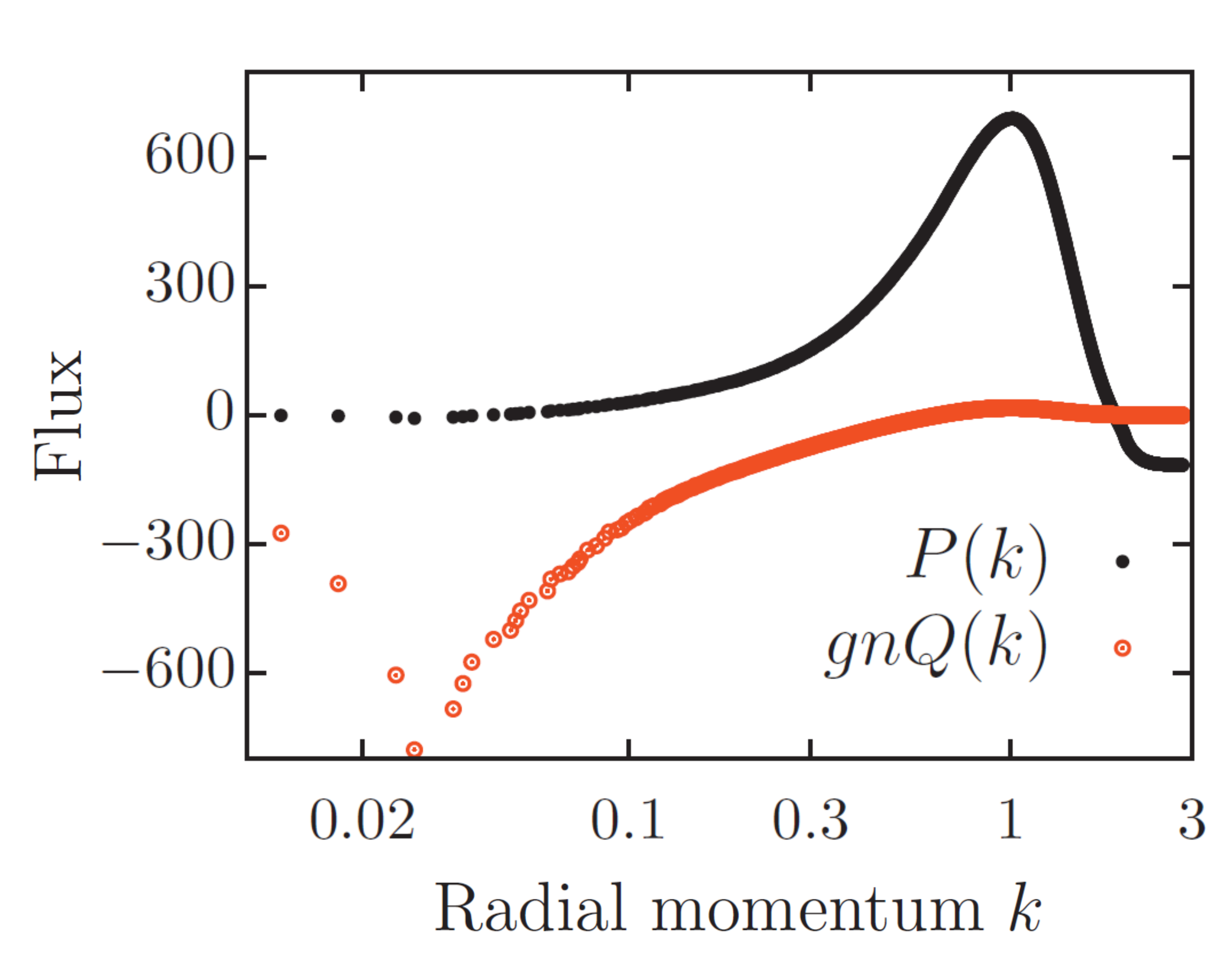}
 \caption{Direct kinetic-energy and inverse particle fluxes, $P(k)$ and $Q(k)$, at an evolution time where a bimodal momentum distribution has emerged, see Ref.~\cite{Nowak:2011sk}. 
 Note the logarithmic $k$-axis. 
 A positive kinetic-energy flux is seen in the UV, a negative particle flux in the IR. 
 Figure taken from Ref.~\cite{Nowak:2013juc}.
}
 \label{fig:PFlux2D3D}
 \end{figure}

\begin{figure*}[t]
\center
\includegraphics[width=0.78\textwidth]{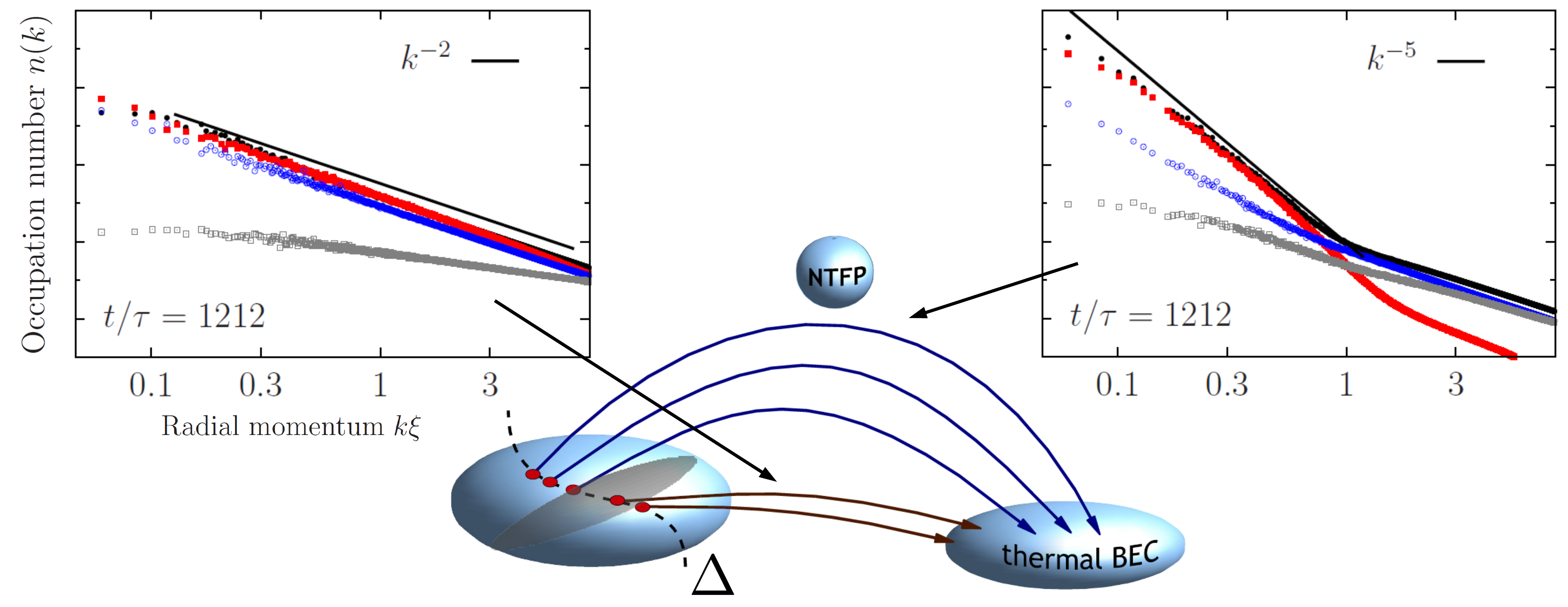}
\caption{Depending on the strength $\Delta$ of the initial cooling quench, where $\Delta$ characterizes the initial high momentum decay $k^{-\Delta}$, a Bose gas in $d=3$ dimensions can thermalize in a near-adiabatic manner to a Bose-Einstein condensate.
Alternatively, it can first approach a non-thermal fixed point (NTFP). Near the fixed point the evolution is critically slowed down and the spectrum is characterized by a steep IR power law, $n(k)\sim k^{-5}$.
Such scaling behavior is found for strong cooling quenches, $\Delta \gtrsim 3$.
Furthermore, dynamical scale separation of the compressible (blue points) and incompressible (red points) components of the velocity field of the gas (cf.~Ref.~\cite{Nowak:2011sk}) occurs.
Note that the incompressible component shows the transverse vortical flow.
The quantum-pressure component is depicted by the grey points.
The radial momentum $k$ is given in units of the healing length $\xi = [2 m g \rho]^{-1/2}$, with mean density $\rho$, and the time $t$ in units of $\tau = m \xi^2$. Note the double-logarithmic scale.
Figure adapted from Ref.~\cite{Nowak:2013juc}.
}
\label{fig:NTFPScheme}
\end{figure*}

\subsection{Defect dominated fixed points}
\label{sec:DefectNTFP}
To illustrate the scaling behavior of a vortex gas at a non-thermal fixed point, we consider the time evolution of an isolated two-dimensional Bose gas whose initial field configuration is chosen such that the condensate density in position space varies between zero and some maximum value.
This can be achieved by macroscopically populating a few of the lowest momentum modes of the system.
Vortices are then created within shock waves forming during the non-linear evolution of the coherent matter-wave field \cite{Nowak:2010tm}.
Alternatively, one can populate momentum modes up to a maximum scale $k_{\mathrm{q}}$, with the corresponding phases in each mode chosen randomly (so-called box initial condition) which also leads to the creation of vortices \cite{Berloff2002a}.
The approach of a non-thermal fixed point is marked by a self-similarly diluting ensemble of vortices and anti-vortices.
From a turbulence point of view, the scaling behavior is characterized by an inverse cascade of particle excitations, possibly accompanied by a direct energy cascade towards the UV. 

During the time evolution the system runs through different stages, see Ref.~\cite{Nowak:2013juc} for details.
On short time scales the dynamics is driven by scattering between the macroscopically occupied modes.
In the next stage of the evolution, one observes strong phase and density gradients forming due to the non-linear evolution.
Those phase gradients lead to the formation of vortices and anti-vortices.
During the stage of unbinding those vortex--anti-vortex pairs and diluting the defects, the evolution slows down and the correlations evolve self-similarly.

A bimodal distribution emerges characterizing the approach of the non-thermal fixed point, cf.~Fig.~\ref{fig:NTFP}.
A UV exponent $\zeta_P^{\mathrm{UV}} \simeq d = 2$ indicates the weak-wave-turbulence prediction in Eq.~(\ref{eq:ZetaUV}) to apply.
In the IR, a steep power-law exponent  $\zeta^{\mathrm{IR}}_Q \simeq d+2 = 4$ is found, which agrees well with with the prediction for the Porod tail from the theory of phase ordering kinetics \cite{Bray1994a.AdvPhys.43.357}.
The power law arises from the algebraic fall-off of the superfluid velocity $|\mathbf{v}|\sim1/r$ with distance $r$ from the core of a single vortex.
The Porod law also indicates correlations in the distances between the defects which in the above case are randomly distributed.
It appears at momenta larger than the mean inverse distance between vortices and anti-vortices and smaller than the inverse core size.

At late times, after the last vortical excitations have disappeared, the entire spectrum becomes thermal, i.e., exhibits the standard  Rayleigh-Jeans scaling, $n(k)\sim T/k^{2}$. 
Note that, in $d=2$ dimensions, the exponent describing weak wave turbulence in the UV  is identical to the exponent in the Rayleigh-Jeans regime. 
Signs of a weak-wave-turbulence exponent of $\zeta_P^{\mathrm{UV}} \simeq d = 3$ have been observed when performing the simulation in $d=3$ dimensions \cite{Nowak:2011sk}.

In the vicinity of the non-thermal fixed point, the system picks the exponents  $\zeta_P^{\mathrm{UV}} \simeq d$ and $\zeta^{\mathrm{IR}}_Q \simeq d+2$ due to the fluxes underlying the stationary but non-equilibrium
distributions. 
Studying the radial particle and energy flux distributions $Q_k$ and $P_k$, see Fig.~\ref{fig:PFlux2D3D}, on time scales within the scaling regime defined by the bimodal structure of the momentum distribution, one finds that the transport process can be interpreted in terms of an inverse particle transport in the IR and a direct energy transport in the UV.
Note that the non-zero energy flux in the UV underpins a weak-wave-turbulence cascade while, by the exponent $\zeta$ it is indistinguishable from thermal scaling.
At late times, thermalization causes the kinetic-energy flux $P$ to almost vanish.
However, $Q$ still reshuffles particles and therefore energy, with the zero mode acting as a sink, keeping the system out of equilibrium close to the non-thermal fixed point.

In the numerical simulations discussed above, the system was initialized in a specific configuration that caused the approach of a non-thermal fixed point.
In the following, we will address the question of the relevance of the strength with which the system is driven away from thermal equilibrium for the approach of the fixed point. 
The corresponding numerical simulations were performed in $d=3$ dimensions \cite{Nowak:2012gd}. 
We parametrize the initial field in momentum space, $\Phi({\mathbf{k}}, 0) = \sqrt{n({\mathbf{k}}, 0)} \exp\{i\varphi({\mathbf{k}}, 0)\}$, in terms of a randomly chosen phase $\varphi({\mathbf{k}}, 0) \in [0, 2\pi)$ and a density $n({\mathbf{k}}, 0) = f (k) \nu_{\mathbf{k}}$, with $\nu_{\mathbf{k}} \geq 0$. 
For each momentum $\mathbf{k}$, the $\nu_{\mathbf{k}}$ are drawn from an exponential distribution $P(\nu_{\mathbf{k}}) = \exp(-\nu_{\mathbf{k}})$.
The resulting occupation number spectrum is flat at low $k$ and falls off according to the function $f (k) = f_{\Delta} / (k_0^{\Delta} + k^{\Delta})$.
The parameter $\Delta$ controls the deviation from a thermal decay with $\Delta = 2$. 
$k_0$ is a momentum cutoff and $f_{\Delta}$ the normalization.
The generated initial configurations are directly overpopulated momentum distributions.

\begin{figure}[t]
\begin{center}
\includegraphics{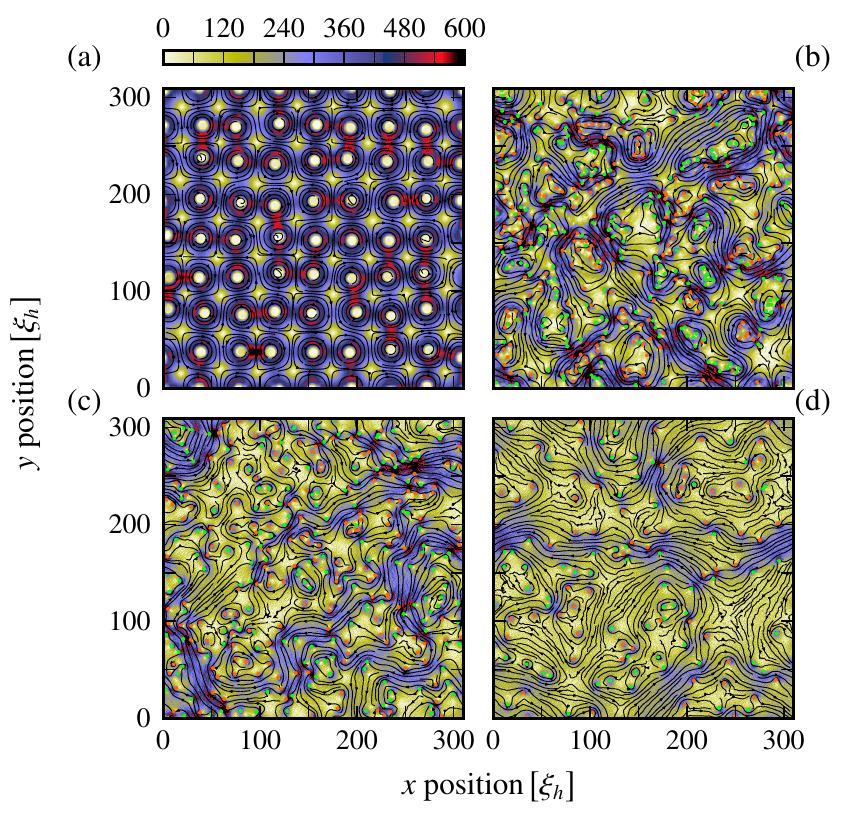}
\caption{Snapshots of the time evolving hydrodynamic velocity field $\mathbf{v} = \nabla \mathrm{arg} ( \Phi) /m$ of a 2D Bose gas in a square 
  volume with periodic boundary conditions, starting from a
  lattice of non-elementary vortices with alternating winding numbers $w=\pm6$,
  arranged in a checker-board manner
  (panel (a)).
  Color encodes the modulus of the velocity $\lvert \mathbf{v} \rvert$.
  The orientation of the velocity field is indicated by the black flow lines.
Panels (b)--(d) show snapshots at times
  $t=\left\{300,10^3,10^4\right\}\,\xi_h^2$, with healing length scale $\xi_h = 1/\sqrt{2mg\rho}$. 
  The initial vortices quickly break up into clusters of elementary vortices and anti-vortices with $w=\pm1$ which are marked by the orange and green dots, respectively.
  A strong vortex clustering is present in the early non-universal stage of the evolution (panel (b)).
  It leads to strong coherent flows during the later stages shown exemplarily in panels (c) and (d), where the vortices and 
  anti-vortices mutually annihilate in a strongly anomalous manner. 
Figure taken from Ref.~\cite{Karl2017b.NJP19.093014}.
  }
   \label{fig:vorlat}
   \end{center}
\end{figure}

\begin{figure}[t]
\centering
    \includegraphics[width= 0.94\columnwidth]{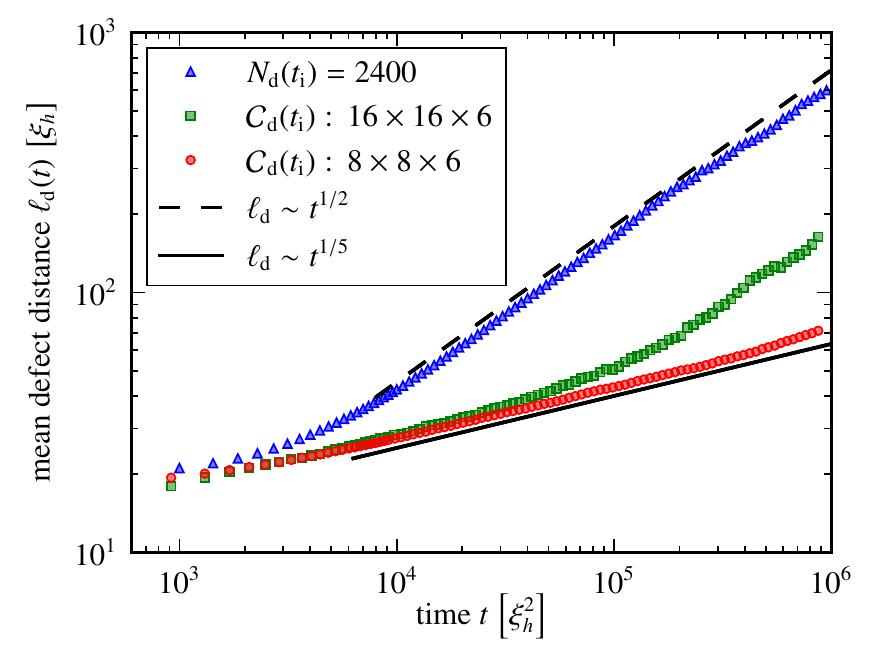}
    \caption{Mean defect distance $\ell_\text{d}$ as function of time,
      starting from different initial vortex configurations at time $t_{\text i} =0$  .
      The blue triangles depict the evolution from a random distribution of
      $N_{\text{d}}(t_{\text i}) = 2400$ elementary vortices and anti-vortices.
      The green squares and red circles correspond to the evolution from 
      an irregular square lattice of $16\times16$ and $8\times8$ non-elementary
      vortices with winding numbers $w=\pm6$ as in Fig.~\ref{fig:vorlat}.
      The growth of the mean defect distance
      is well described by power laws $\ell_\text{d}(t)\sim t^{\,\beta_{\text d}}$ 
      with $\beta_{\text d}={1/5}$ (anomalous fixed point, solid line) and $\beta_{\text d}={1/2}$ (Gaussian fixed point, dashed line), respectively. 
      Depending on the initial condition chosen, the system can also
      first approach the anomalous fixed point before it shows faster scaling reminiscent of
      the Gaussian fixed point (see data marked by green squares). Units as in Fig.~\ref{fig:vorlat}. 
      Note the double-logarithmic scale.
      Figure taken from Ref.~\cite{Karl2017b.NJP19.093014}.
     }
  \label{fig:VortexNumber}
\end{figure}

At sufficiently late evolution times, the occupation number spectra developing from different initial $\Delta$ differ strongly, see Fig.~\ref{fig:NTFPScheme}.
For $\Delta \gtrsim 3$, the system approaches a non-thermal fixed point, characterized by a bimodal structure of the spectra, with a power-law behavior $n(k) \sim k^{-5}$ in the IR and $n(k) \sim k^{-2}$ in the UV.
The bimodal structure decays towards a global $n(k) \sim k^{-2}$ at very long times (not shown). 
For $\Delta \lesssim 3$, the momentum distribution goes over directly to thermal Rayleigh-Jeans scaling $n(k) \sim T /k^2$.
The larger the quench strength $\Delta$, the closer the system approaches the non-thermal fixed point, see Ref.~\cite{Nowak:2012gd} for details. 

The numerical simulations indicate that cutting away sufficiently much population at high momenta initially is necessary if the system is supposed to approach the non-thermal fixed point. 
Hence, only strong cooling quenches allow for the build-up of a steep population far into the IR and a bimodal scaling evolution, see also the  discussion in Sec.~\ref{sec:ScalingKinEq}.

So far, we have discussed the scaling exponents characterizing the power-law scaling of the momentum distribution, $n(k) \sim k^{-\zeta}$, at a non-thermal fixed point in a vortex gas. 
Depending on the initial momentum distribution, the system is either attracted to a non-thermal fixed point or simply relaxes back to thermal equilibrium.
In general, more than one attractor can exist for the dynamical evolution of the system.
Consequently, different types of universal evolution with different power laws for each attractor are found.
Which  type  of  evolution  is realized depends on the macroscopic properties of the initial
state and on the stability properties of the attractors only.

Preparing  far-from-equilibrium states by imprinting phase defects, i.e., quantum  vortex  excitations,  into  an  otherwise
strongly phase-coherent two-dimensional  Bose condensate, the approach of two different non-thermal fixed points can be triggered \cite{Karl2017b.NJP19.093014}.
This initial state leads to coherent hydrodynamic propagation  of  vortices  on  the  background  of  the  otherwise phase-coherent gas, with little sound excitations present, which plays a key role for the observation of scaling.
Different kinds of initial states are realized by varying the number of defects, their arrangement, and their winding numbers.
A strongly anomalous non-thermal fixed point as well as a standard dissipative fixed point related to coarsening according to the Hohenberg-Halperin model A \cite{Hohenberg1977a,Bray1994a.AdvPhys.43.357} have been identified in numerical simulations \cite{Karl2017b.NJP19.093014}.

\begin{figure*}[t]
\includegraphics[width=0.9\textwidth]{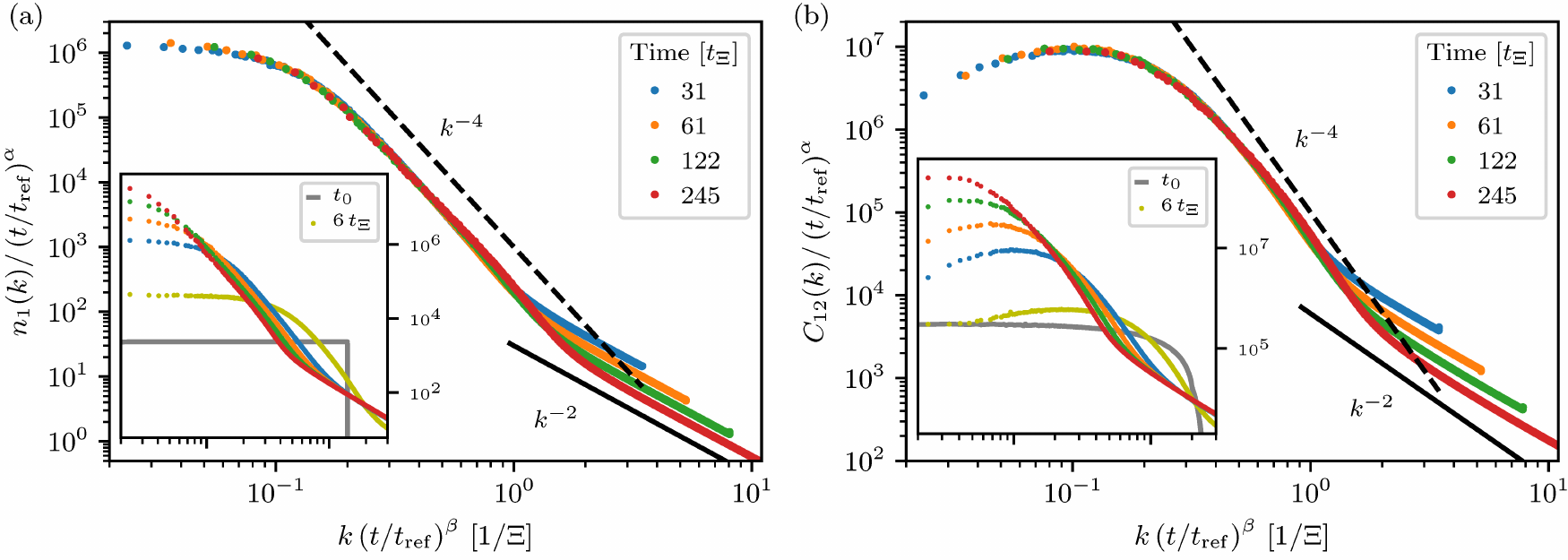}
\caption{(a) Universal scaling of the occupation number $n_{1}(k)\equiv n_{1}(\mathbf{k},t)$ near a non-thermal fixed point, according to Eq.~(\ref{eq:ScalingHypothesis}). 
The inset shows the evolution starting from a `box' momentum distribution $n_{1}(\mathbf{k},t_{0})=n_{0}\,\Theta(k_\mathrm{q}-|\mathbf{k}|)$ (grey line), with $n_{0}=(4\pi k_{\mathrm{q}}^{3})^{-1}\rho^{(0)}$ and $k_\mathrm{q}=1.4\,k_{\Xi}$, at five different evolution times (colored dots).
Here, $\rho^{(0)}$ is the constant mean background density and $k_{\Xi} = \sqrt{2mg\rho^{(0)}}$ marks the momentum scale associated with the corresponding healing length of the system.
The collapse of the data onto a universal scaling function, with reference time $t_\mathrm{ref}= 31 \, t_{\Xi}$, shows the scaling in space and time.
Within the time window $t_{\mathrm{ref}}=200\,t_{\Xi} \leq t \leq 350\,t_{\Xi}$, we extract the scaling exponents $\alpha=1.62\pm0.37$, $\beta=0.53\pm0.09$.
(b) Universal scaling dynamics of the correlator measuring the spatial fluctuations of the relative phases, $C_{12}(k,t) = \langle\lvert (\Phi_1^{\dagger} \Phi_2)(k,t)\rvert ^2\rangle $ for the same system. 
The collapse of the data onto a universal function, which has a similar shape as for $n_{1}$, shows that the scaling 
behavior at the non-thermal fixed point is driven by relative  phase fluctuations.
Within the same time window as stated in (a), we extract the scaling exponents $\alpha=1.48\pm0.18$, $\beta=0.51\pm0.06$.
The scaling exponents are in good agreement with the analytical predictions \cite{Mikheev:2018adp,Chantesana:2018qsb.PhysRevA.99.043620} of $\beta = 1/2$ and $\alpha = d \beta = 3/2$.
The evolution time is measured in units of $t_\Xi^{-1} = g \rho^{(0)} / 2\pi$, the momentum in terms of the inverse healing-length scale $\Xi^{-1} = k_{\Xi}$.
Figure taken from Ref.~\cite{Schmied:2018upn.PhysRevLett.122.170404}.
\label{fig:NTFPEvolution} 
}
\end{figure*}

The anomalous fixed point is approached if the coupling of the defects to the background sound fluctuations is sufficiently suppressed.
Starting from a lattice of  non-elementary vortices with alternating winding numbers $w = \pm 6$ arranged in a checker-board manner, the vortices were found to arrange within clusters of elementary defects with either positive or negative winding, $w = \pm 1$, such that the formation of closely bound vortex--anti-vortex dipoles is suppressed, see Fig.~\ref{fig:vorlat} for snapshots of the corresponding velocity field. 
The clustering leads to a steep power law scaling $n(k) \sim k^{-5.7}$, i.e., an exponent $\zeta \simeq 5.7$.
The subsequent scaling evolution is driven by the  mutual annihilation of vortices and anti-vortices that proceeds in a strongly anomalous manner.

The defect dilution at the anomalous fixed point is much slower than in the standard dissipative case. 
The anomalous fixed point is characterized by a universal scaling exponent $\beta = 1/(2-\eta) \simeq 1/5$ that governs the self-similar evolution of the momentum distribution, in the IR regime of momenta, according to 
\begin{equation}
\label{eq:ScalingHypothesis}
n(k, t) = (t/t_\mathrm{ref})^{\alpha} f ( [t/t_\mathrm{ref}]^ \beta k),
\end{equation}
with universal scaling function $f$ and some reference time $t_{\mathrm{ref}}$ within the temporal scaling regime.
Due to particle number conservation within the regime of  low momenta and times considered in the numerical simulations the scaling exponents are related according to Eq.~(\ref{eq:ConservedQPDens}) such that $\alpha = d \beta \simeq 2/5$.
The large anomalous exponent $\eta \simeq -3$ is related to a large dynamical exponent $z = 2 - \eta \simeq5$.
The observed strongly slowed scaling can be interpreted as being due to mutual defect annihilation following three-vortex collisions and has recently been found consistent with experimental data \cite{Simula2014a.PhysRevLett.113.165302,Groszek2016a.PhysRevA.93.043614, Johnstone2019a.Science.364.1267}.


In contrast to the above scenario, starting from a random spatial distribution of elementary defects, with equal number of vortices and anti-vortices, the system approaches the standard dissipative fixed point characterized by the scaling exponents $\beta \simeq 1/2$ and $\eta \simeq 0$.
Particle number conservation leads to $\alpha = d \beta \simeq 1$.
Due to the vanishing anomalous exponent this fixed point was referred to as the (near) Gaussian non-thermal fixed point, while the numerical findings are also compatible with a small but non-zero anomalous exponent \cite{Karl2017b.NJP19.093014,Schachner:2016frd}.
The observed scaling is associated with the mutual annihilation of elementary vortices and anti-vortices randomly distributed on the phase-coherent background. 
The Porod exponent $\zeta \simeq 4$ is consistent with a dilute ensemble of randomly distributed vortices in $d = 2$
dimensions \cite{Bray1994a.AdvPhys.43.357,Nowak:2011sk,Schole:2012kt}. 

The distinctly different scaling behavior, emerging from the two initial vortex configurations chosen, becomes clearly visible in the time evolution of the mean defect distance $\ell_\text{d}$, see Fig.~\ref{fig:VortexNumber}.  
As the mean defect distance sets the characteristic IR length scale of the system, its scaling evolution is described by the IR scaling exponent $\beta$ according to $\ell_\text{d} (t) \sim t^{\beta}$. 
Interestingly, depending on the number of non-elementary vortices in the initial configuration, the system can first approach the anomalous fixed point and subsequently show faster scaling reminiscent of the Gaussian fixed point.
A similar behavior has been observed in experiment \cite{Johnstone2019a.Science.364.1267}.
A transition between different scalings has also been found in a relativistic $\phi^{6}$ model with attractive quartic interactions \cite{Berges:2017ldx}.

\begin{figure*}[t]
\includegraphics[width=0.9\textwidth]{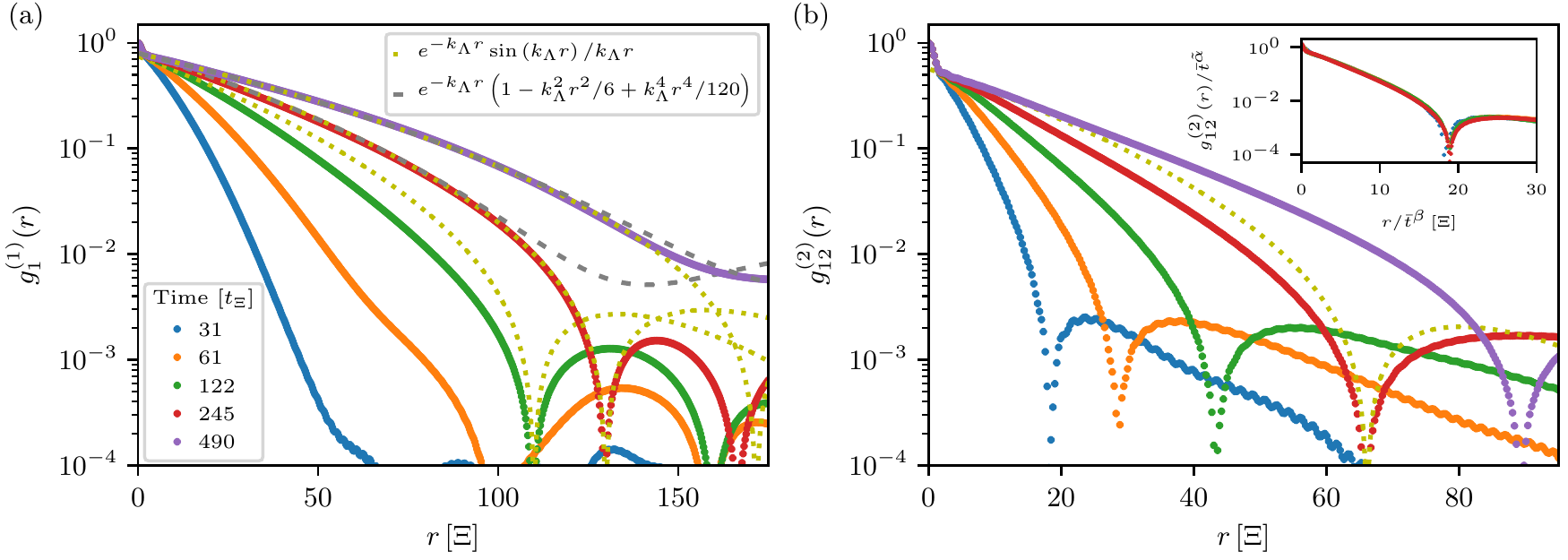}
\caption{(a) Time evolution of the first-order coherence function $g^{(1)}_{1}(r)=g^{(1)}_{1}(\mathbf{r},t) = \langle\Phi_{1}^{\dag}(\mathbf{x}+\mathbf{r},t)\Phi_{1}(\mathbf{x},t)\rangle$ showing violations of universal scaling at larger distances (five different times, colored dots).
The scaling form (\ref{eq:NTFPscalingg1}), with inverse coherence length scale $k_{\Lambda}(t)$ fitted to match the first zero of the sine, is depicted by the yellow dotted lines.
The polynomial approximation of the sinc as given in the legend is visualized by the grey dashed lines. 
At later times, the numerical results agree well with the single-scale function up to the first zero in $r$.
At the latest time shown finite-size effects already become relevant. 
(b) Corresponding second-order coherence function $g_{12}^{(2)} (r,t)=\langle\Phi_{1}^{\dag}(\mathbf{x}+\mathbf{r},t)\Phi_{2}(\mathbf{x}+\mathbf{r},t)\Phi_{2}^{\dag}(\mathbf{x},t)\Phi_{1}(\mathbf{x},t)\rangle$  measuring the spatial fluctuations of the relative phases between components 1 and 2 (colored dots, same times as in (a)).
The inset shows the rescaled correlation function $\bar{t}^{-\tilde\alpha}g_{12}^{(2)}(\bar{t}^{-\beta}r,t_\mathrm{ref})$, with $\beta = 0.6$, $\tilde{\alpha} = -0.15$, and $\bar{t} = t/t_{\mathrm{ref}}$. 
The collapse of the data onto a single function, for short distances $r$, indicates that the scaling violations are considerably weaker than for $g^{(1)}_{1}$.
Here, $t_{\mathrm{ref}}= 31 \, t_{\Xi}$ denotes the reference time. 
The evolution time is measured in units of $t_\Xi^{-1} = g \rho^{(0)} / 2\pi$, the distance $r$ is given in terms of the healing-length scale $\Xi= 1/\sqrt{2mg\rho^{(0)}}$, with constant mean background density $\rho^{(0)}$.
Figure taken from Ref.~\cite{Schmied:2018upn.PhysRevLett.122.170404}. 
\label{fig:RelativePhase12}}
\end{figure*}

\subsection{Fluctuation dominated fixed points}
\label{sec:FluctuationNTFP}
If topological defects are subdominant, the scaling behavior of a Bose gas at a non-thermal fixed point can be 
different.
To illustrate this, we consider an $(N=3)$-component dilute Bose gas in $d=3$ dimensions, quenched far out of equilibrium \cite{Schmied:2018upn.PhysRevLett.122.170404,Mikheev:2018adp}.
The system is described by an $O(3)\times U(1)$ 
or $U(3)$ symmetric Gross-Pitaevskii model with quartic contact interaction in the total density, see Eq.~(\ref{eq:HamiltonianMulticomponentBoseGas}).

As for the $d=3$ vortex case above, the initial far-from-equilibrium  state at time $t_0$ is given by a `box' momentum distribution $n_a(\mathbf{k},t_0) = n_0 \Theta(k_\mathrm{q} - \lvert \mathbf{k} \rvert$), which is constant up to some cutoff scale $k_\mathrm{q}$. 
The initial phase angles $\varphi_{a}(\mathbf{k},t_{0})$ of the Bose fields in Fourier space, $\Phi_{a}(\mathbf{k},t_{0})=\sqrt{n_{0}}\exp[i\varphi_{a}(\mathbf{k},t_{0})]$, are chosen randomly on the circle and thus uncorrelated \footnote{Note that the phase angles of the fundamental fields $\Phi_{a}(\mathbf{k},t)$ in Fourier space are different from the Fourier transforms of the phase angles in position space, $\varphi_{a}(\mathbf{k},t)\not=\theta_{a}(\mathbf{k},t)$.}.
Such an initial condition can be realized by means of a strong cooling quench, cf.~Sec.~\ref{sec:NTFP}.
After a few collision times, the system shows universal scaling indicating the approach of the non-thermal fixed point.

The time evolution of the occupation number $n_1(k)$ as well as of the correlator measuring the spatial fluctuations of the relative phases, $C_{12}(k,t) = \langle\lvert (\Phi_1^{\dagger} \Phi_2)(k,t)\rvert ^2\rangle $, is depicted in Fig.~\ref{fig:NTFPEvolution}.
For both observables we obtain a collapse of the data onto a universal scaling function.
This collapse shows the universal scaling in space and time according to the scaling hypothesis in Eq.~(\ref{eq:ScalingHypothesis}).
Within the time window $t_{\mathrm{ref}}=200\,t_{\Xi} \leq t \leq 350\,t_{\Xi}$,
the extracted scaling exponents are $\alpha=1.62\pm0.37$, $\beta=0.53\pm0.09$ for the scaling of the occupation number $n_1$ and, respectively, $\alpha=1.48\pm0.18$, $\beta=0.51\pm0.06$ for $C_{12}$.
The scaling exponents agree well with the analytical predictions of $\beta = 1/2$ and $\alpha = d \beta = 3/2$ \cite{Mikheev:2018adp,Chantesana:2018qsb.PhysRevA.99.043620}.
The scaling collapse of the relative phase correlator $C_{12}$ clearly shows that the scaling behavior at the non-thermal fixed point is driven by relative phase fluctuations.
The spatial scaling exponent $\zeta \simeq 4$ characterizing the occupation number distribution $n_1(k) \sim k^{-\zeta}$ also confirms analytical predictions neglecting defects \cite{Chantesana:2018qsb.PhysRevA.99.043620}.

The physical picture is that the $O(3)\times U(1)$ symmetric interactions suppress total density fluctuations while allowing the densities of the separate components to be shuffled around freely provided that the total density stays constant.
In this way, also relative phase fluctuations can occur, which reflect the counter-motion of particles and correspond to strongly excited Goldstone modes.

\section{Prescaling}
\label{sec:Prescaling}
So far, we have discussed the scaling behavior of dilute Bose gases \emph{at} a non-thermal fixed point. 
It remains, though, an unresolved question how precisely quantum many-body systems evolve from a given initial state to such a fixed point.
As a typical feature of this evolution towards the fixed point we propose prescaling \cite{Schmied:2018upn.PhysRevLett.122.170404}, see also the illustration in the right panel of Fig.~\ref{fig:Prethermalization}. 

Prescaling \cite{Wetterich2018a.privcomm}, motivated by the concept of partial fixed points \cite{Bonini1999a}, means that certain correlation functions, already at comparatively early times and short distances, scale with the universal exponents predicted for the fixed point. 
The fixed point itself will only be reached much later in time and, in a finite-size system, may not be reached at all.

During the stage of prescaling, weak scaling violations occur for the correlations at larger distances.
Such scaling violations only slowly vanish.
In fact, it turns out that they affect not only the scaling exponents but in particular also the shape of the associated scaling functions.
The existence of prescaling has been proposed on the basis of numerical simulations of an $(N=3)$-component dilute Bose gas in $d=3$ dimensions, quenched far out of equilibrium \cite{Schmied:2018upn.PhysRevLett.122.170404}.
The initial far-from-equilibrium state at time $t_0$ is given by  the configuration presented in Sec.~\ref{sec:FluctuationNTFP}.

Scaling behavior at a non-thermal fixed point is commonly extracted from momentum-space correlators \cite{Orioli:2015dxa,Walz:2017ffj}.
Prescaling, however, is more easily seen in position-space correlations.
An intuitive choice, based on the momentum-space treatments, is to study  the first-order spatial coherence function $g^{(1)}_{a}(\mathbf{r},t)=\langle\Phi_{a}^{\dag}(\mathbf{x}+\mathbf{r},t)\Phi_{a}(\mathbf{x},t)\rangle$, see Fig.~\ref{fig:RelativePhase12}a.
For long evolution times it is found to approach the exponential $\times$ cardinal-sine form   
\begin{equation}
  g_{a}^{(1)}(\mathbf{r},t) \approx \rho_{a}^{(0)}e^{-k_{\Lambda}(t)\,|\mathbf{r}|}
  \,{\mathrm{sinc}\big(k_{\Lambda}(t)\,|\mathbf{r}|\big)}\,.
 \label{eq:NTFPscalingg1}
\end{equation}
While the particle density $\rho_{a}^{(0)}$ is uniform, the phase oscillates and fluctuates on a scale given by the inverse coherence length $k_{\Lambda}$. 
At the fixed point, this length scale rescales in time according to $k_{\Lambda}(t)\sim t^{-\beta}$, with universal scaling exponent $\beta$.
Note that the form of the first-order coherence function, Eq.~(\ref{eq:NTFPscalingg1}), differs from a pure exponential obtained analytically within a Gaussian approximation of the relation  between the phase-angle and the phase correlators, see Ref.~\cite{Mikheev:2018adp}
for details.

As fluctuations of local density differences, in contrast to the fluctuations of the total density, are not suppressed, Goldstone excitations of the relative phases can become relevant.
An observable sensitive to the relative phases $\theta_{a}-\theta_{b}$ is the second-order coherence function   $g^{(2)}_{ab}(\mathbf{r},t)=\langle\Phi_{a}^{\dag}(\mathbf{x}+\mathbf{r},t)\Phi_{b}(\mathbf{x}+\mathbf{r},t)\Phi_{b}^{\dag}(\mathbf{x},t)\Phi_{a}(\mathbf{x},t)\rangle$.
The time evolution of this correlation function, see Fig.~\ref{fig:RelativePhase12}b, reveals weaker scaling violations than obtained for $g_1^{(1)}$, indicating that the fixed point scaling is driven by relative phase fluctuations.

A temporal scaling analysis of the correlation functions $g^{(1)}_{1}(\mathbf{r},t)$ and $g^{(2)}_{12}(\mathbf{r},t)$ provides a direct way to extract the scaling exponent $\beta$.
If, however, the fixed-point scaling is not fully developed, the time evolution of the correlations is not described by a single scale $k_\Lambda (t)$. 
During prescaling we expect approximate scaling to emerge on short distances and to subsequently spread towards longer distances.
To be independent of the particular form of the scaling function we make use of  a general polynomial fit of the form $g (\mathbf{r},t)\simeq c_{0} + c_1 k_{\Lambda,1}(t) \, r+c_{2}[k_{\Lambda,2}(t) \, r]^{2}+ c_3[k_{\Lambda,3}(t) \, r]^{3} + c_{4}[k_{\Lambda,4}(t) \, r]^{4}+\mathcal{O}(r^{5})\}$, at short distances $r$, to study the scaling behavior of the different types of correlations.
Note that the fit is applied to distances $r\gtrsim 5\,\Xi$ avoiding the short-distance thermal peak present in the correlation functions.

\begin{figure}[t]
\centering
\includegraphics[width=0.97\columnwidth]{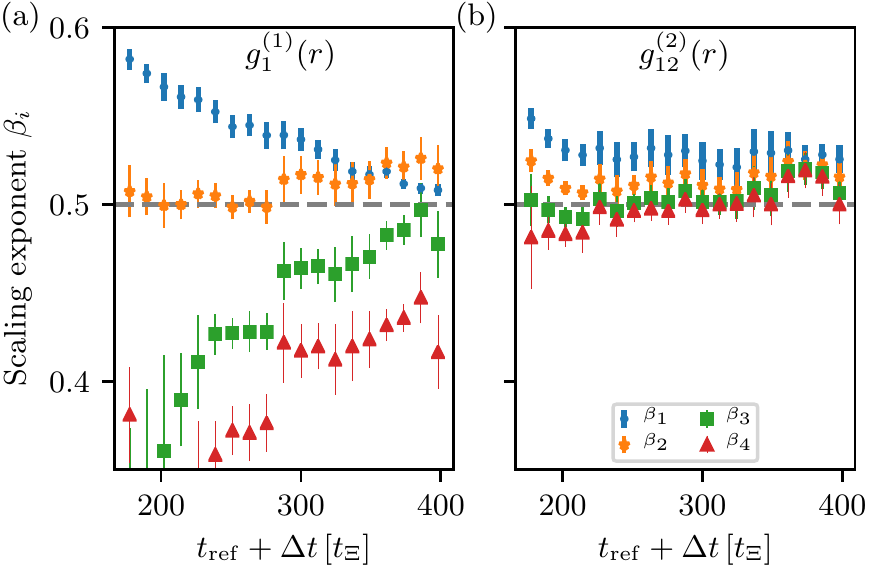}
\caption{Prescaling of position-space correlations.
(a)  Scaling exponents $\beta_{i}$ characterizing the time evolution of the inverse coherence length scale $k_{\Lambda,i}(t)\sim t^{-\beta_{i}}$ for $i =1,2,3, 4$. 
The $k_{\Lambda,i}(t)$ are extracted by means of a polynomial fit of the first-order coherence function $g^{(1)}_{1}(r,t)$, shown in Fig.~\ref{fig:RelativePhase12}a, up to order $r^{4}$ at small distances $r$. The index $i$ corresponds to the associated order of the polynomial. See main text for details.
 (b) Scaling exponents $\beta_{i}$ 
 obtained from an analogous polynomial fit of $g^{(2)}_{12}(r,t)$, see Fig.~\ref{fig:RelativePhase12}b.
 Prescaling is seen by the scaling exponents $\beta_i$ settling in to, within errors, constant values for the lower orders of the polynomial fit.
While $g_{1}^{(1)}(r)$ exhibits scaling violations, $g_{12}^{(2)}(r)$ already shows scaling up to order $r^{4}$.
The scaling exponent $\beta_i \simeq 0.5$, extracted for times $t_{\mathrm{ref}} + \Delta t \gtrsim250\,t_{\Xi}$, is in good agreement with the predicted scaling exponent $\beta=1/2$.
The $\beta_{i}$ are averaged over the time window $[t_\mathrm{ref}, t_\mathrm{ref} + \Delta t]$ with $\Delta t = 146\,t_{\Xi}$ as well as over a set of fits with different fit ranges. 
Errors are given by the standard deviation of the exponents of the set.
Units as in Fig.~\ref{fig:RelativePhase12}.
Figure taken from Ref.~\cite{Schmied:2018upn.PhysRevLett.122.170404}.
\label{fig:WindowFitAllkLi}
}
\end{figure}

The scaling exponents $\beta_{i}$, associated with the order $i$ of the polynomial fit, are obtained by taking the logarithmic derivative of $k_{\Lambda,i}(t)$ with respect to the time $t$ and averaging it over a fixed time window $\Delta t$.
The resulting exponents $\beta_{i}$ for both, $g_{1}^{(1)}$ and $g^{(2)}_{12}$, for $i=1,2,3,4$ are depicted in Fig.~\ref{fig:WindowFitAllkLi}.
We remark that the exponents shown are additionally averaged over a set of fits with different fit ranges to account for fluctuations arising from the choice of a particular fit range.

Prescaling is quantitatively seen by the scaling exponents settling in to stationary values for the lower orders of the polynomial expansion. 
However, scaling in the higher orders is not yet fully developed for the times considered in the numerical simulations. 
This leads to the scaling violations observed in Fig.~\ref{fig:RelativePhase12}. 
Comparing Figs.~\ref{fig:WindowFitAllkLi}a and b, we find that different observables can enter the stage of prescaling on different time scales.
Hence, establishing the full scaling function and the associated scaling exponents is, to some degree, observable-dependent.

The value $\beta_{i}\simeq0.5$ found at late evolution times for the scaling of $k_{\Lambda,i}(t) \sim t^{-\beta_i}$, for $i =1,2, 3, 4 $, parameterizing $g_{12}^{(2)}$, and for $i=1,2$ in the case of $g^{(1)}_{1}$, is in good agreement with the analytically predicted value $\beta=1/2$ by means of the LEEFT, see Sec.~\ref{sec:LEEFT}.
As the LEEFT approach covers the limiting cases of $N =1$ and $N \to \infty$, the numerical results suggest that the universality class does not depend on $N$.
This reflects that the $U(N)$ symmetry is broken during prescaling while the $U(1)$ symmetries are still intact as long as no condensate is present.

\section{Outlook}
\label{sec:Outlook}
In this article we discussed the concept of non-thermal fixed points on the basis of a dilute Bose gas.
We outlined a kinetic theory as well as a low-energy effective field theory approach which allowed for analytical predictions of the scaling behavior at the non-thermal fixed point.
While the scaling evolution of the fundamental fields is considered within non-perturbative kinetic theory, the low-energy effective field theory describes the dynamics and scaling of phase excitations in the system based on a perturbative approximation in the non-linearities.
We presented a variety of numerical studies corroborating the analytical predictions.

By treating the dilute Bose gas within a low-energy effective field theory, it was possible to predict scaling exponents, characterizing the time evolution of the system at a non-thermal fixed point, in the limiting cases of $N=1$ and $N \to \infty$. 
A rigorous approach to analytically study the scaling evolution at intermediate $N$ is missing so far. 
In addition, the Luttinger-liquid based description neglects defects of all kinds in the system.
It is an interesting pathway for the future to derive a low-energy effective theory in presence of defects.

Based on the analytical treatments as well as numerical studies it can be concluded that universal dynamics at non-thermal fixed points can emerge in rather different manners depending on the properties of the systems.
On the one hand, the dilution of defects can drive the scaling evolution leading to strong wave turbulence in the IR regime of momenta.
If defects are subdominant in the system, relative phase excitations can play a crucial role for the observed self-similar evolution.

Tuning the initial condition of the system enabled to identify the key features leading to the approach of a non-thermal fixed point.
Whether the system shows self-similar universal scaling dynamics or directly relaxes to thermal equilibrium crucially depends on the strength of cooling quenches applied to the system.

We furthermore discussed the possibility that a system can be attracted to more than one fixed point. 
In the case of a dilute Bose gas in two spatial dimensions, the initial vortex configuration played the key role whether the system exhibits strongly anomalous or standard diffusive universal scaling. 

Conducting numerical simulations of a three-component Bose gas in three spatial dimensions, the existence of prescaling as a feature of the evolution towards the non-thermal fixed point has been proposed. 
As the system prescales on comparatively short times scales it can be studied in present-day experimental systems.

In this brief overview we focussed on $N$-component Bose gases characterized by $O(N)$-symmetric interactions. 
Multicomponent spinor Bose gases break the $O(N)$ symmetry of the models discussed here due to the presence of spin-spin interactions and a possible quadratic Zeeman energy shift arising from external magnetic fields.
Spinor Bose gases are extremely well controlled in present-day experiments. 
Non-equilibrium dynamics following quantum quenches has been investigated in several experimental systems \cite{Higbie2005a,Sadler2006a,Guzman2011a}. 
Recently, universal scaling with exponent $\beta \simeq 0.5$ has been observed experimentally in a spin-1 Bose gas confined in a quasi one-dimensional trapping geometry \cite{Prufer:2018hto}.
Realizing experimental observations of the system for evolution times up to several seconds gives the opportunity to address universal scaling dynamics in such systems. 
Numerical simulations of a one-dimensional spin-1 Bose gas revealed a scaling exponent of $\beta \simeq 0.25$. While the purely one-dimensional dynamics of the spin-1 Bose gas is driven by defects populating the system, the experimentally observed scaling is different in nature as defects are absent in the quasi one-dimensional setting.
Including interactions that break the $O(N)$ symmetry of the models into the analytical approaches presented in this article is part of ongoing research.

\section*{Acknowledgements}
{The authors thank I.~Aliaga Sirvent, J.~Berges, K.~Boguslavski, R.~B\"uck\-er, I.~Chantesana, S.~Diehl, S.~Erne, F.~Essler, 
M.~Karl, P.~Kunkel, S.~Lannig, D.~Linnemann, A.~Mazeliauskas, B.~Nowak, M.~K.~Oberthaler, J.~M.~Pawlowski, A.~Pi{\~n}eiro Orioli, M.~Pr\"ufer, R.~F.~Rosa-Medina Pimentel, J.~Schmiedmayer, J.~Scho\-le, T.~Schr\"oder, H.~Strobel, and C.~Wetterich for discussions and collaborations on the topics described here, in particular A.~Pi{\~n}eiro Orioli and K.~Boguslavski for their careful reading of the final manuscript.
This overview article has been written for the proceedings of the \emph{Julian Schwinger Centennial Conference and Workshop} held in Singapore in February 2018.
T.G.~thanks the Julian Schwinger Foundation for Physics Research for support and the Institute of Advanced Studies at Nanyang Technological University, Singapore, for its hospitality.
Original work summarized here was supported by the Horizon-2020 programme of the EU (AQuS, No. 640800; ERC Adv.~Grant EntangleGen, Project-ID 694561),  by DFG (GA677/8 and SFB 1225 ISOQUANT) and by Heidelberg University (CQD).

\bibliographystyle{apsrev4-1}

%


\end{document}